\documentclass[11pt,letterpaper]{iopart}

\usepackage{iopams,amsfonts,amssymb,amsthm} 

\usepackage{etex}
\usepackage{cite}
\usepackage{latexsym}
\usepackage{rawfonts}
\usepackage{graphicx}
\usepackage[usenames,dvipsnames,svgnames]{xcolor}
\usepackage{bbm}
\usepackage{latexsym}
\usepackage{multirow}
\usepackage{rotating}
\usepackage{lscape}
\usepackage{graphicx} 
\usepackage[format=plain,labelfont=it,textfont=it]{caption}
\usepackage[format=plain,labelfont=it,textfont=it]{subcaption}
\usepackage{xcolor}
\usepackage{fancybox}
\usepackage{ifmtarg}
\usepackage{fancyhdr}
\usepackage{wrapfig}
\usepackage{hyperref}
\usepackage{tikz}
\usepackage{ulem}
\usepackage[flushleft]{threeparttable}

\input prepictex
\input pictex
\input postpictex

\def\2;{\;\;}

\def\eps{\epsilon}
\def\P#1#2{{P^{(#1)}_{#2}}}
\def\Pr#1{{\mathbb{P}^{(#1)}}}
\def\pr{{\prime}}

\def\IntZ{{\mathbb Z}}
\def\NatNu{{\mathbb N}}

\def\RealN{{\mathbb R}}

\def\Ref#1{(\ref{#1})}

\def\z#1{{z_#1}}

\def\binom#1#2{{{#1}\choose{#2}}}
\def\Bi#1#2{{\binom{#1}{#2}}}

\def\Sfrac#1#2{\hbox{\large $\frac{#1}{#2}$}}
\def\sfrac#1#2{\hbox{\nor $\frac{#1}{#2}$}}

\def\LB{\left(}         \def\RB{\right)}

\def\lfl{\!\left\lfloor} \def\rfl{\right\rfloor\!}
\def\LC{\left\{}       \def\RC{\right\}}

\def\nor{\normalsize}

\def\vv{{\;\hbox{\Large $|$}\;}}
\def\svv{{\;\hbox{$|$}\;}}

\def\edge#1#2{{\langle #1 \hspace{0.85pt}{\sim}\hspace{0.85pt} #2 \rangle}}

\def\plus{{\hspace{0.85pt}{+}\hspace{0.85pt}}}
\def\minus{{\hspace{0.85pt}{-}\hspace{0.85pt}}}

\def\tiny#1{\scalebox{0.75}{#1}}

\definecolor{blue}{rgb}{0,0.18,0.39}
\definecolor{RoyalBlue}{rgb}{0,0.2,0.7}

\definecolor{Maroon}{cmyk}{0,0.87,0.68,0.62}
\definecolor{Brown}{rgb}{0.7,0.3,0}
\definecolor{Navy}{rgb}{0.3,0.0,0.4}
\definecolor{Red}{cmyk}{0,1,1,0}
\definecolor{BrickRed}{cmyk}{0.16,0.89,0.61,0.02}
\definecolor{DarkRed}{cmyk}{0,1,1,0.5}
\definecolor{DarkBlue}{cmyk}{1,1,0,0.2}
\definecolor{DarkGreen}{cmyk}{1,0,1,0.4}
\definecolor{Green}{cmyk}{1,0,1,0}
\definecolor{DarkBrown}{cmyk}{0,0.81,1,0.6}
\definecolor{OrangeRed}{cmyk}{0,1,0.87,0}
\definecolor{RedOrange}{cmyk}{0,0.77,0.87,0}
\definecolor{Orange}{cmyk}{0,0.61,0.87,0}
\definecolor{Offwhite}{rgb}{.8,0.9,.8}
\definecolor{Offwhite2}{cmyk}{.04,.02,.01,0}
\definecolor{Tan}{rgb}{0.82,0.70,0.55}
\definecolor{Blue}{rgb}{0,0,1}
\definecolor{RoyalBlue}{rgb}{0.25,0.41,0.88}
\definecolor{Sepia}{rgb}{0.37,0.14,0.07}
\definecolor{myblue}{cmyk}{0.025,0.05,0,0}
\definecolor{Mahogany}{cmyk}{0.18,0.87,1,0.08}

\definecolor{green1}{cmyk}{0.25,0,0.76,0}
\definecolor{green2}{cmyk}{0.25,0,0.76,0.07}
\definecolor{green3}{cmyk}{0.25,0,0.76,0.20}
\definecolor{green4}{cmyk}{0.25,0,0.75,0.30}
\definecolor{green5}{cmyk}{0.25,0,0.75,0.40}
\definecolor{green6}{cmyk}{0.25,0,0.75,0.50}

\definecolor{B02}{cmyk}{0,0.14,0.22,0.12}
\definecolor{B03}{cmyk}{0,0.16,0.26,0.16}
\definecolor{B04}{cmyk}{0,0.19,0.28,0.19}
\definecolor{B05}{cmyk}{0,0.25,0.32,0.25}
\definecolor{B06}{cmyk}{0,0.31,0.36,0.31}
\definecolor{B07}{cmyk}{0,0.37,0.40,0.37}
\definecolor{B08}{cmyk}{0,0.46,0.46,0.46}
\definecolor{B09}{cmyk}{0,0.55,0.52,0.54}
\definecolor{B10}{cmyk}{0,0.69,0.61,0.62}
\definecolor{B11}{cmyk}{0,0.78,0.70,0.68}
\definecolor{B12}{cmyk}{0,0.93,0.85,0.60}
\definecolor{B13}{cmyk}{0.25,1,0.6,0.50}
\definecolor{B14}{cmyk}{0.5,1,0.30,0.40}
\definecolor{B15}{cmyk}{0.75,1,0,0.30}

\definecolor{C02}{cmyk}{0,0.22,0.14,0.12}
\definecolor{C03}{cmyk}{0,0.26,0.16,0.16}
\definecolor{C04}{cmyk}{0,0.28,0.19,0.19}
\definecolor{C05}{cmyk}{0,0.32,0.25,0.25}
\definecolor{C06}{cmyk}{0,0.36,0.31,0.31}
\definecolor{C07}{cmyk}{0,0.40,0.37,0.37}
\definecolor{C08}{cmyk}{0,0.46,0.46,0.46}
\definecolor{C09}{cmyk}{0,0.52,0.55,0.54}
\definecolor{C10}{cmyk}{0,0.61,0.69,0.62}
\definecolor{C11}{cmyk}{0,0.70,0.78,0.68}
\definecolor{C12}{cmyk}{0,0.85,0.93,0.60}
\definecolor{C13}{cmyk}{0.25,0.60,1,0.50}
\definecolor{C14}{cmyk}{0.5,0.30,1,0.40}
\definecolor{C15}{cmyk}{0.75,0,1,0.30}

\def\dps{\displaystyle}
\newcommand*{\erf}[0]{\hbox{erf}}

\begin{document}

\title[Trajectories of directed lattice paths]{
Trajectories of directed lattice paths}

\author{EJ Janse van Rensburg$^2\ddagger$}
\address{\sf$^2$Department of Mathematics and Statistics, 
York University, Toronto, Ontario M3J~1P3, Canada\\}
\ead{$\ddagger$\href{mailto:rensburg@yorku.ca}{rensburg@yorku.ca}}
\vspace{10pt}
\begin{indented}
\item[]\today
\end{indented}

\begin{abstract}
The distribution of monomers along a linear polymer grafted on a hard wall is 
modelled by determining the probability distribution of occupied vertices of 
Dyck path and Dyck meander models of adsorbing linear polymers.  For 
example, the probability that a Dyck path passes through the lattice site 
with coordinates $(\lfl \eps n \rfl,\lfl \delta \sqrt{n}\rfl)$ in the square lattice, 
for $0 < \eps < 1$ and $\delta\geq 0$, is determined asymptotically as $n\to\infty$ and
this uncovers the probability density of vertices along Dyck paths in the limit as the length
of the path $n$ approaches infinity:
\[ \Pr{D} (\eps,\delta) = \frac{4\delta^2}{\sqrt{\pi\,\eps^3(1\minus\eps)^3}}
\, e^{-\delta^2/\eps(1\minus \eps)} . \]
The properties of a polymer coating of a hard wall and the density or distribution of monomers 
in the coating is relevant in applications such as the stabilisation of a colloid dispersion by 
a polymer or in a drug delivery system such as a drug-eluding stent covered by a grafted polymer.  
\end{abstract}

%
\vspace{2pc}


%
%

\section{Introduction}
\label{s1}

\begin{figure}[t!]
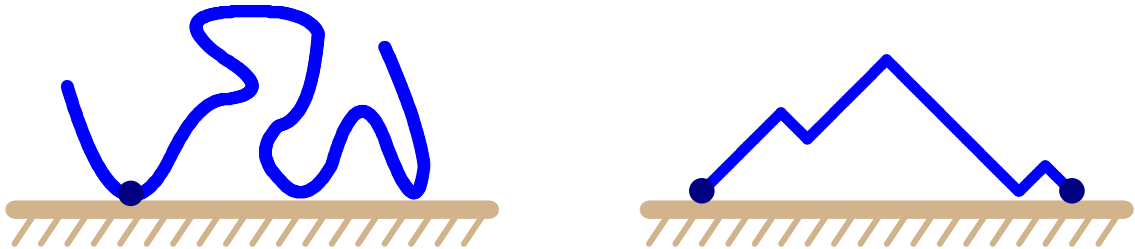

\beginpicture
\setcoordinatesystem units <1pt,1pt>
\setplotarea x from -30 to 250, y from -20 to 80

\color{Tan}
\setplotsymbol ({\Large$\bullet$})
\plot -20 -7 160 -7 /
\setplotsymbol ({\LARGE$\cdot$})
\plot -20 -20 -10 -5 / \plot -10 -20 0 -5 / \plot 0 -20 10 -5 / \plot 10 -20 20 -5 /
\plot 20 -20 30 -5 / \plot 30 -20 40 -5 / \plot 40 -20 50 -5 / \plot 50 -20 60 -5 /
\plot 60 -20 70 -5 / \plot 70 -20 80 -5 / \plot 80 -20 90 -5 / \plot 90 -20 100 -5 /
\plot 100 -20 110 -5 / \plot 110 -20 120 -5 / \plot 120 -20 130 -5 / \plot 130 -20 140 -5 /
\plot 140 -20 150 -5 / \plot 150 -20 160 -5 /

\color{Blue}
\setplotsymbol ({$\bullet$})
\setquadratic
\plot 0 40 20 0 40 15  50 30 60 35  70 40 60 50 50 65 80 68
90 65 95 60 90 35 80 25  75 15 80 5 90 0 100 10
110 30 120 20 130 0 135 10  130 30 120 55  /
\setlinear
\color{NavyBlue}
\multiput {\huge$\bullet$} at 24 -1 /

\setcoordinatesystem units <1pt,1pt> point at -240 0
\setplotarea x from -30 to 250, y from -20 to 80

\color{Tan}
\setplotsymbol ({\Large$\bullet$})
\plot -20 -7 160 -7 /
\setplotsymbol ({\LARGE$\cdot$})
\plot -20 -20 -10 -5 / \plot -10 -20 0 -5 / \plot 0 -20 10 -5 / \plot 10 -20 20 -5 /
\plot 20 -20 30 -5 / \plot 30 -20 40 -5 / \plot 40 -20 50 -5 / \plot 50 -20 60 -5 /
\plot 60 -20 70 -5 / \plot 70 -20 80 -5 / \plot 80 -20 90 -5 / \plot 90 -20 100 -5 /
\plot 100 -20 110 -5 / \plot 110 -20 120 -5 / \plot 120 -20 130 -5 / \plot 130 -20 140 -5 /
\plot 140 -20 150 -5 / \plot 150 -20 160 -5 /

\color{Blue}
\setplotsymbol ({\footnotesize$\bullet$})
\plot 0 0  10 10  20 20  30 30 40 20 50 30 60 40 70 50 80 40
90 30 100 20 110 10 120 0 130 10 140 0  /
\color{NavyBlue}
\multiput {\huge$\bullet$} at 0 0 140 0 /

\color{Black}
\normalcolor
\endpicture
\caption{A schematic diagram of a linear polymer grafted on a surface (left).
A Dyck path model of a grafted linear polymer (right).  The Dyck path steps
North-East or South-East in the square lattice, and it is grafted into
the surface at its first and last nodes.}
\label{f1}
\end{figure}

The physical properties of a polymer coating on surfaces or on suspended
particles have been the subject of research for many decades.  For example, 
linear polymers grafted on colloid particles or on other surfaces are 
important in biomedical and industrial applications including in the steric 
stabilisation of colloid dispersions \cite{FS93,N83}, or the use of polymers in drug
delivery systems \cite{S06,JM12,LKSP10}.  Examples of these drug delivery 
systems involve nanoparticle-polymer systems \cite{TOS01,SYS15},
or polymer coatings on drug-eluding stents \cite{FHS01}.  In these
systems a polymer is adsorbed or grafted onto the surface (see figure
\ref{f1} for a schematic diagram) and it coils from the surface to form 
a polymer layer in which the drug absorbs and is stabilised until it is delivered 
at a target site.  The properties of the polymeric coating is of significant importance.  
If it is hydrophylic then it extends away from the surface forming a thicker 
and less dense coating. If, on the other hand, it is hydrophobic, then it forms 
a thinner and denser layer.  Both these states may be important in 
applications, and from a physical perspective, the density profile of the 
polymeric coating may give indications of its suitability  in an application. 

In this paper the density profile of a grafted polymer is modelled using directed
lattice paths.  Directed models of adsorbing and grafted polymers were pioneered 
in references \cite{CP88,PFF88,F90,F91,W98}and are useful in modelling entropic 
and scaling properties of linear polymers.  The simplest of these models is 
shown in figure \ref{f1} and is a \textit{Dyck path}.  Dyck paths are directed 
lattice paths from the origin giving North-East and South-East steps in the 
upper half square lattice and are conditioned to end in the $x$-axis.  The path 
visits the $x$-axis, and if these visits are weighed by a parameter $a$,
then for large $a>0$ the path adsorbs in the $x$-axis.  This is a model of an
adsorbing grafted linear polymer; see references \cite{BY95,BEO98,BORW05,RJvR04,JvR05}.
Dyck path models of linear polymers are exactly solvable, and these models
are still examined for both their mathematical properties, and as models of 
polymer entropy and phase behaviour \cite{JvR10A,IJvR12,JvRP13}.

\begin{figure}[t!]
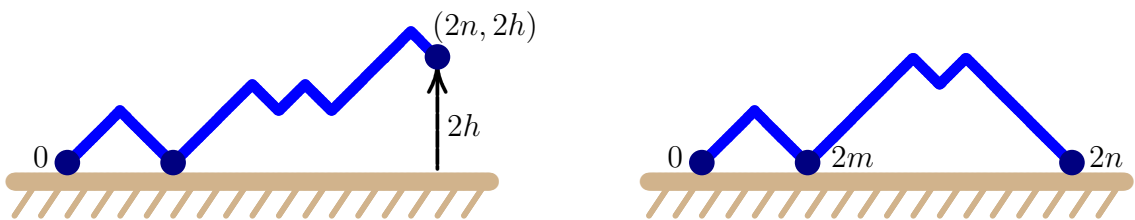

\beginpicture
\setcoordinatesystem units <1pt,1pt>
\setplotarea x from -30 to 250, y from -20 to 60

\setplotsymbol ({$\cdot$})
\arrow <10pt> [.2,.67] from 140 -2 to 140 37
\put {$0$} at -10 3
\put {$2h$} at 150 15
\put {$(2n,2h)$} at 158 52

\color{Tan}
\setplotsymbol ({\Large$\bullet$})
\plot -20 -7 160 -7 /
\setplotsymbol ({\LARGE$\cdot$})
\plot -20 -20 -10 -5 / \plot -10 -20 0 -5 / \plot 0 -20 10 -5 / \plot 10 -20 20 -5 /
\plot 20 -20 30 -5 / \plot 30 -20 40 -5 / \plot 40 -20 50 -5 / \plot 50 -20 60 -5 /
\plot 60 -20 70 -5 / \plot 70 -20 80 -5 / \plot 80 -20 90 -5 / \plot 90 -20 100 -5 /
\plot 100 -20 110 -5 / \plot 110 -20 120 -5 / \plot 120 -20 130 -5 / \plot 130 -20 140 -5 /
\plot 140 -20 150 -5 / \plot 150 -20 160 -5 /

\color{Blue}
\setplotsymbol ({\footnotesize$\bullet$})
\plot 0 0  10 10  20 20  30 10 40 0 50 10 60 20 70 30 80 20
90 30 100 20 110 30 120 40 130 50 140 40  /
\color{NavyBlue}
\multiput {\huge$\bullet$} at 0 0 40 0 140 40 /

\color{Black}
\normalcolor

\setcoordinatesystem units <1pt,1pt> point at -240 0
\setplotarea x from -30 to 250, y from -20 to 60

\put {$0$} at -10 3
\put {$2m$} at 57 3
\put {$2n$} at 153 3

\color{Tan}
\setplotsymbol ({\Large$\bullet$})
\plot -20 -7 160 -7 /
\setplotsymbol ({\LARGE$\cdot$})
\plot -20 -20 -10 -5 / \plot -10 -20 0 -5 / \plot 0 -20 10 -5 / \plot 10 -20 20 -5 /
\plot 20 -20 30 -5 / \plot 30 -20 40 -5 / \plot 40 -20 50 -5 / \plot 50 -20 60 -5 /
\plot 60 -20 70 -5 / \plot 70 -20 80 -5 / \plot 80 -20 90 -5 / \plot 90 -20 100 -5 /
\plot 100 -20 110 -5 / \plot 110 -20 120 -5 / \plot 120 -20 130 -5 / \plot 130 -20 140 -5 /
\plot 140 -20 150 -5 / \plot 150 -20 160 -5 /

\color{Blue}
\setplotsymbol ({\footnotesize$\bullet$})
\plot 0 0  10 10  20 20  30 10 40 0 50 10 60 20 70 30 80 40
90 30 100 40 110 30 120 20 130 10 140 0  /
\color{NavyBlue}
\multiput {\huge$\bullet$} at 0 0 40 0 140 0 /

\color{Black}
\normalcolor
\endpicture
\caption{(Left) A Dyck meander from the origin to the point $(2n,2h)$.  The path 
steps in the half square lattice $\IntZ_+^2$ from the origin, has length $2n$ steps,
and the height of its last node is $2h$.  (Right) A Dyck path of length $2n$
from the origin, and passing through the vertex $(2m,0)$.}
\label{f2}
\end{figure}

The \textit{half square lattice} is defined by 
$\IntZ_+^2 = \{(n,m)\svv \hbox{$n,m\in \IntZ$ and $m \geq 0$}\}$.
The boundary of the half square lattice is $\partial\IntZ_+^2$ and it 
is also called the \textit{hard wall} or \textit{adsorbing wall}.
A \textit{directed path} from the origin in $\IntZ_+^2$ is a sequence of
steps (edges) $(\edge{u_i}{u_{i+1}})$ such that $u_0=(0,0)$,
$u_{i+1}=u_i+(1,{\pm}1)$.  If a vertex $u=(m,h)$ then $h$ is the
\textit{height} of $u$.  If the last vertex $u_n = (n,h)$, then the path
has \textit{length} $n$.  That is, we consider a model of directed paths
giving North-East or South-East steps in the half square lattice.

If a directed path in $\IntZ_+^2$ from the origin has length $2n$, and its 
last vertex has coordinates $u_{2n}=(2n,0)$, then it is a \textit{Dyck path}
(and it last vertex has height zero). 

A path from the origin and ending in a vertex at arbitrary height $h$
is a \textit{Dyck meander}.  We shall abuse this terminology by simply
calling them ``meanders''.

 A Dyck path is grafted at its first and last vertices
in the hard wall or adsorbing wall, while a meander is grafted to the wall only
at its first vertex.  These models are illustrated in figure \ref{f2}, and they are
models of adsorbing grafted linear polymers.

The number of meanders of length $2n$  and ending in the vertex
$(2n,2h)$ is 
\begin{equation}
D(2n,2h) = \frac{2h+1}{n+h+1} \Bi{2n}{n+h} .
\label{1}
\end{equation}
Putting $h=0$ gives the number of Dyck paths of length $2n$:
\begin{equation}
D(2n,0) = \frac{1}{n+1} \Bi{2n}{n} .
\label{2}
\end{equation}
Note that $\dps D(2n,0) = C_n$ where the $C_n$ are the Catalan numbers.

The vertices in a directed path are distributed in the area on and above the wall.
For a given $n\in\NatNu$, there is a probability that a directed path will pass
through a vertex of coordinates $(2n,2h)$.  This is a (discrete) probability measure
on the square lattice, which is zero outside the first quadrant and on vertices
of odd parity (a vertex $(n,m)$ has odd parity if $n+m$ is odd).

\begin{figure}[h!]
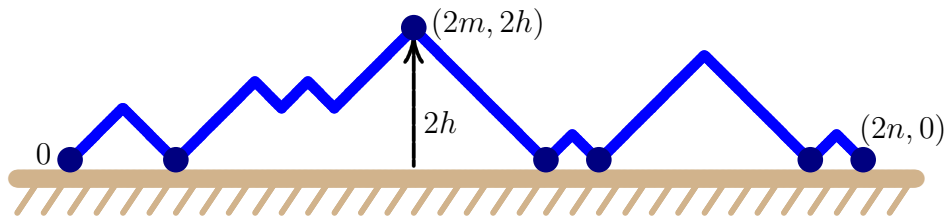

\beginpicture
\setcoordinatesystem units <1pt,1pt>
\setplotarea x from -70 to 250, y from -20 to 60

\setplotsymbol ({$\cdot$})
\arrow <10pt> [.2,.67] from 130 -2 to 130 47
\put {$0$} at -10 3
\put {$2h$} at 140 15
\put {$(2m,2h)$} at 158 52
\put {$(2n,0)$} at 315 12

\color{Tan}
\setplotsymbol ({\Large$\bullet$})
\plot -20 -7 320 -7 /
\setplotsymbol ({\LARGE$\cdot$})
\plot -20 -20 -10 -5 / \plot -10 -20 0 -5 / \plot 0 -20 10 -5 / \plot 10 -20 20 -5 /
\plot 20 -20 30 -5 / \plot 30 -20 40 -5 / \plot 40 -20 50 -5 / \plot 50 -20 60 -5 /
\plot 60 -20 70 -5 / \plot 70 -20 80 -5 / \plot 80 -20 90 -5 / \plot 90 -20 100 -5 /
\plot 100 -20 110 -5 / \plot 110 -20 120 -5 / \plot 120 -20 130 -5 / \plot 130 -20 140 -5 /
\plot 140 -20 150 -5 / \plot 150 -20 160 -5 /

\plot 160 -20 170 -5 / \plot 170 -20 180 -5 / \plot 180 -20 190 -5 / \plot 190 -20 200 -5 /
\plot 200 -20 210 -5 / \plot 210 -20 220 -5 / \plot 220 -20 230 -5 / \plot 230 -20 240 -5 /
\plot 240 -20 250 -5 / \plot 250 -20 260 -5 / \plot 260 -20 270 -5 / \plot 270 -20 280 -5 /
\plot 280 -20 290 -5 / \plot 290 -20 300 -5 / \plot 300 -20 310 -5 / \plot 310 -20 320 -5 /

\color{Blue}
\setplotsymbol ({\footnotesize$\bullet$})
\plot 0 0  10 10  20 20  30 10 40 0 50 10 60 20 70 30 80 20
90 30 100 20 110 30 120 40 130 50 /
\plot 130 50 140 40 150 30 160 20 170 10 180 0 190 10 200 0 210 10
220 20 230 30 240 40 250 30 260 20 270 10 280 0 290 10 300 0 / 
\color{NavyBlue}
\multiput {\huge$\bullet$} at 0 0 40 0 130 50 180 0 200 0 280 0 300 0  /

\color{Black}
\normalcolor
\endpicture
\caption{A Dyck path of length $2n$ passing through the vertex $(2m,2h)$
is composed of two meanders (one reversed).  The first meander is
from the origin to the vertex $(2m,2h)$, and the second starts in the vertex
$(2n,0)$ and then steps in the North-West and South-West directions until it
terminates in $(2m,2h)$.}
\label{f3}
\end{figure}

\begin{figure}[t]
\begin{center}
\includegraphics[width=0.475\textwidth]{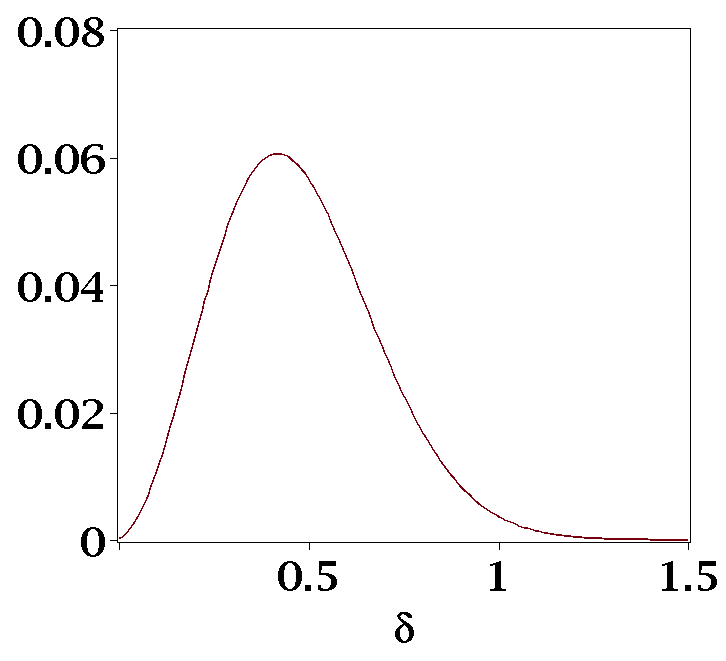}
\includegraphics[width=0.475\textwidth]{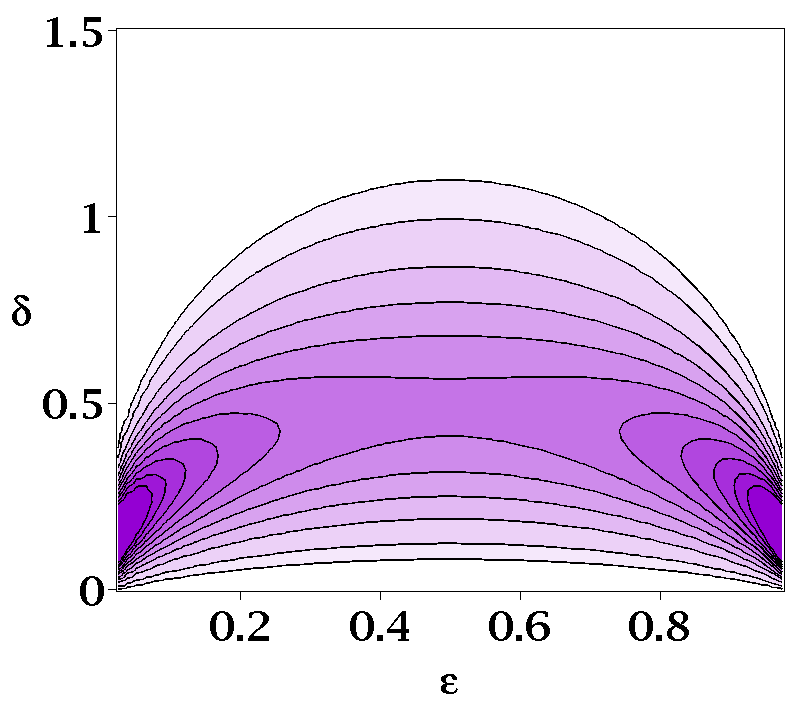}
\end{center}
\caption{(Left) The probability that a Dyck path passes through a vertex
of height $2\lfl\delta \sqrt{n}\rfl$ a fractional distance $\eps=1/4$ along its
total length $2n=2000$.  
(Right)  A contour plot of the probability density
$\P{D}{2n} (2\lfl \eps n\rfl,2\lfl\delta\sqrt{n}\rfl)$ (equation \Ref{3}) for Dyck paths of
total length $2n=2000$.
}
\label{f4}
\end{figure}

The probability that a Dyck path of length $2n$ passes through the site with
coordinates $(2m,2h)$ (see figure \ref{f3}) is given by the ratio of the number of
Dyck paths passing through $(2m,2h)$ divided by the number of Dyck paths of
length $2n$: 
\begin{equation}
\P{D}{2n}(2m,2h) = \frac{D(2m,2h).D(2n-2m,2h)}{D(2n,0)} .
\label{3}
\end{equation}
It can be checked that $\sum_{h\geq 0} \P{D}{2n}(2m,2h) = 1$.  
The probability $\P{D}{2n}(2m,2h)$ is conditioned on all the paths
having the same weight, and are therefore uniformly sampled in this 
model.  That is, $\P{D}{2n}(2m,2h)$ is the probability that a randomly 
sampled path passes through the point with coordinates $(2m,2h)$. The natural
scaling underlying this model is $O(n)$ in the horizontal direction and
$O(\sqrt{n})$ in the vertical direction, since the directed path is a random
walk in the vertical direction.  Thus, introduce the scaling $m=\lfl \eps n\rfl$ 
and $h=\lfl\delta\sqrt{n}\rfl$ in equation \Ref{3}.  Plotting
$\P{D}{2n}(2\lfl \eps n\rfl,2\lfl\delta\sqrt{n}\rfl)$ for $2n=2000$ against $\delta$
for $\eps=0.25$ gives the distribution in figure \ref{f4}(left).  This shows a peak 
well away from the hard wall, indicating that the path tends to drift away
from the wall. 

In the right panel of figure \ref{f4} a contour plot of 
$\P{D}{2n}(2\lfl \eps n\rfl,2\lfl\delta\sqrt{n}\rfl)$ for $2n=2000$ is shown in the
$\eps\delta$-plane (with $0<\eps<1$ and $\delta>0$. The parameter 
$\epsilon$ is the fractional distance along the $x$-axis, while
$\delta$ is the rescaled height of vertices in the path (such that
$\delta\approx h/\sqrt{n}$).  This shows that in probability, the path
drifts away from the hard wall, and only returns at its endpoint, visiting
the hard wall infrequently.

Taking the limit $n\to\infty$ (with scaling $m=\lfl \eps n\rfl$ and 
$h=\lfl\delta\sqrt{n}\rfl$) gives the limiting probabability density function
\begin{equation}
\Pr{D}(\eps,\delta) = \lim_{n\to\infty} \P{D}{2n}(2\lfl\eps n\rfl,2\lfl \delta\sqrt{n} \rfl)
= \frac{4\,\delta^2\,e^{-\delta^2/\epsilon(1\minus\epsilon)}
}{\sqrt{\pi\,\epsilon^{3}(1\minus\epsilon)^{3}}}.
\label{4}
\end{equation}
It can be checked that $\int_0^\infty \Pr{D}(\eps,\delta)\,d\delta = 
\int_0^1\int_0^\infty \Pr{D}(\eps,\delta)\,d\delta\,d\epsilon=1$. 
While equation \Ref{4} is exact, its derivation was not done with complete
mathematical rigour. In this paper, our aim is to derive exact probability
densities for the models of Dyck paths, rather than providing proofs that our
results are exact.

\begin{figure}[h!]
\begin{center}
\includegraphics[width=0.675\textwidth]{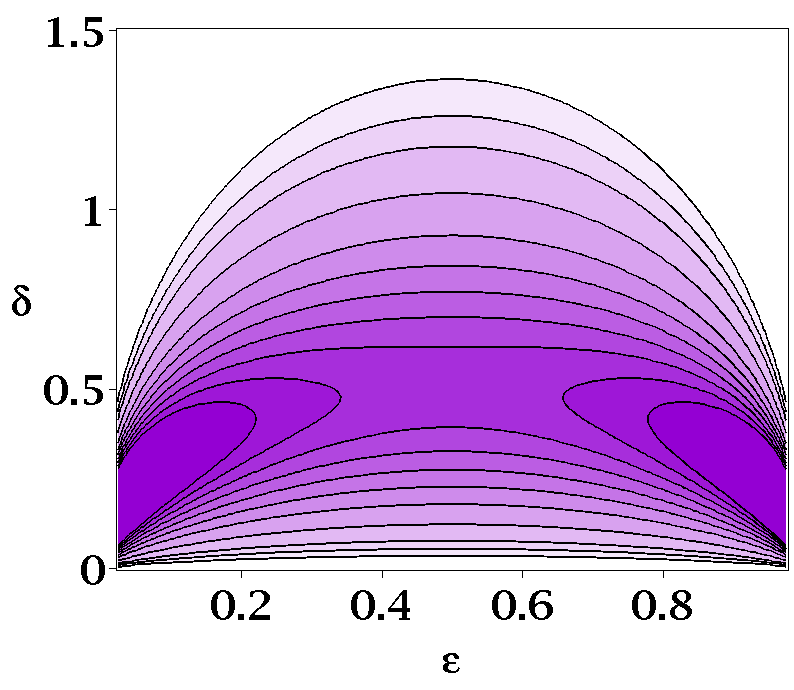}
\end{center}
\caption{The probability density $\Pr{D}(\epsilon,\delta)$ (equation \Ref{4})
of a Dyck path passing through points in the $\epsilon\delta$-plane in the
continuum limit.}
\label{f5}
\end{figure}

The probability density in equation \Ref{4}
induces a probability measure $d\rho(\eps,\delta)$ so that the probability
of a path moving through a subset $A\subseteq \RealN^2$, in the scaling
($n=\infty$) limit is given by
\begin{equation}
\mathbb{P}(A) = \int_A \Pr{D}(\eps,\delta)\,d\lambda 
= \int_A \frac{d\rho(\eps,\delta)}{d\lambda}\,d\lambda 
= \int_A d\rho(\eps,\delta),
\end{equation}
where $\lambda$ is plane measure, and $\Pr{D}(\eps,\delta)$ is the
Radon-Nikodym derivative of $\rho(\eps,\delta)$ with respect to plane measure 
$\lambda$. The probability density $\Pr{D}(\eps,\delta)$ is defined everywhere 
in $\RealN^2$ and is bigger than zero only when $(\eps,\delta)\in [0,1]\times[0,\infty)$.
In figure \ref{f5} a contour plot of $\Pr{D}(\eps,\delta)$ is shown.  Areas in a
darker shade have higher probability densities.

The \textit{mean path} can be calculated by integrating
\begin{equation}
\overline{\delta}(\eps) = \int_0^\infty \delta\, \Pr{D}(\eps,\delta)\,d\delta
= (2/\sqrt{\pi})\sqrt{\eps(1-\eps)} .
\end{equation}

The \textit{modal path} is obtained by determining that value of $\delta$
maximizing $\Pr{D}(\eps,\delta)$.  In this case it is given by
\begin{equation}
\delta_M(\eps) = \sqrt{\eps(1-\eps)}.
\end{equation}
In general, the probability that the path stays underneath the curve 
$c(x) = x\sqrt{\eps(1-\eps)}$ is determined by noting that
\begin{equation}
p(x) = \int_0^1 \int_0^{c(x)} \Pr{D}(\eps,\delta)\,d\delta\,d\eps 
= \erf(x) - (2/\sqrt{\pi})\,e^{-x^2}.
\end{equation}
Putting $x=1$ gives the probability of staying underneat the modal curve, and this
evaluates to $p(1)=0.42759\ldots$, while underneath the mean path
$p(2/\sqrt{\pi}) = 0.53305\ldots$.  In contrast, $c(1/4)\approx 0.01132\ldots$,
and this is a low probability.  Notice that $p(x) \simeq 4\,x^3/3\sqrt{3} + O(x^4)$
so that even for $x=0.1$, $p(x)$ is very small.  These results suggest that there is an
area underneath the mean or modal path with a low probability of being occupied.
This is consistent with the observations that a grafted polymer layer will absorb
molecules as noted earlier in this introductory section, or that a grafted polymer experiences
an entropic repulsion from the hard wall, swelling it away from the hard wall
and creating a low density region next to the hard wall which is shielded by 
higher monomer densities along the modal or mean paths.

In the next sections this paper is organised as follows.  In section \ref{s2.1}
the results above, namely equation \Ref{4} and the results following from it,
are shown.  The model is generalised to Dyck meanders in section \ref{s2.2}.
The results are consistent with those seen for Dyck paths, but the lifted or
free endpoint of the path do give more general probability densities.

In section \ref{s3} a Dyck path model of \textit{adsorbing} linear polymers 
is examined.  Adsorbing Dyck path models have long been studied as a model 
of the adsorption transition in linear polymers grafted to a hard wall
\cite{CP88,PFF88,F90,F91,BY95,W98,RJvR04}.  In this model the paths pass from 
a desorbed phase through a transition into an adsorbed phase.  In the desorbed
phase the paths tend to drift away from the hard wall, while in the adsorbed phase
it tends to stay close to the hard wall.  The critical point separating these phases
is a continuous phase transition, and scaling of the model depends on the 
phase.  The probability density of adsorbed paths is examined in section \ref{s3.1})
and it shows that in the scaling limit the path adheres to the adsorbing wall, not making
any excursions into bulk.  At the critical point (this is also called the 
\textit{special point}) the scaling limit changes and trajectories of the path well away 
from the hard wall may be seen.  This is shown in section \ref{s3.2}.
In the desorbed phase the probability density is identical to that of Dyck Paths 
shown in figure \ref{f5} and equation \Ref{4}, as argued in section \ref{s3.3}.
The results are briefly reviewed in the discussion section \ref{s4}.

\section{Probability densities in models of directed paths}

\subsection{Dyck Paths}
\label{s2.1}

By equations \Ref{1}, \Ref{2} and \Ref{3}, the probability of a Dyck path of 
(even) length $2n$ passing through the lattice site with coordinates $(2m,2h)$ is
\begin{equation}
\hspace{-1cm}
\P{D}{2n}(2m,2h) = \frac{(2h+1)^2(2n+1)}{(m+h+1)(n-m+h+1)}
\scalebox{1.67}{$\frac{\Bi{2m}{m+h}\,\Bi{2(n-m)}{n-m+h}}{\Bi{2n}{n}}$} .
\label{9}
\end{equation}
Proceed by putting $m=\lfl \eps n\rfl$ and $h=\lfl\delta\sqrt{n}\rfl$, and convert
this expression into factorials.  These are approximated asymptotically using the 
approximation derived in equation \Ref{64} in Appendix A:
\begin{equation}
n! = \sqrt{\pi(2n\plus 1/3)}\,n^ne^{-n}\,(1+O(1/n^2)) .
\label{10}
\end{equation}
The asymptotics of $D(2n,0)$ (Catalan numbers) are well known 
\color{red}\cite{V91}\color{black}, and given by
\begin{equation}
D(2n,0) = \frac{1}{\sqrt{\pi n^3}}\,4^n\,(1+O(1/n)) .
\end{equation}

It remains to determine asymptotic expressions for 
$D(2\lfl\sigma n\rfl,2\lfl\delta\sqrt{n}\rfl)$ where $\sigma=\eps$ or 
$\sigma=1\minus\eps$.  Converting the binomial
coefficient into its component factorials, and using the Stirling approximation
in equation \Ref{9} gives, after taking logarithms, followed by expanding
each term and simplifying (using Maple \cite{maple17})
\begin{eqnarray*}
\hspace{-1cm}
\log D(2\lfl\sigma n\rfl,2\lfl\delta\sqrt{n}\rfl)
&= (1\plus 2\,\sigma n)\log 2 + \log \sqrt{\sigma/\pi} + \log (\delta/(\sigma^2 n)) 
- \delta^2/\sigma \\
&\qquad + (1/(2\,\delta)-\delta/\sigma)/\sqrt{n} + O(1/n) .
\end{eqnarray*}
Exponentiating the above, and simplifying, gives
\begin{equation}
\hspace{-1cm}
D(2\lfl\sigma n\rfl,2\lfl\delta\sqrt{n}\rfl)
=\frac{2\,\delta\, 4^{\sigma n}}{\sqrt{\pi\sigma^3}\,n}\;
e^{-(2\,\delta^3\sqrt{n}+2\,\delta^2-\sigma)/(2\,\delta\sigma\sqrt{n})}\times  (1+O(1/n)) .
\label{12}
\end{equation}
Combining these expressions for $D(2n,0)$, 
$D(2\lfl \eps n\rfl,2\lfl \delta\sqrt{n}\rfl)$ and
$D(2\lfl(1\minus\eps)n\rfl,2\lfl\delta\sqrt{n}\rfl)$ approximates the probability:
\begin{equation}
\hspace{-2.5cm}
\P{D}{2n}(2\lfl \eps n\rfl,2\lfl \delta\sqrt{n}\rfl) 
= \frac{4\,\delta^2}{\sqrt{\pi n \eps^3(1\minus\eps)^3}}\,
e^{-(\delta^3\sqrt{n}+\delta^2-\eps(1-\eps))/(\delta \eps(1\minus\eps)\sqrt{n})}
\times (1+O(1/n)) .
\label{13}
\end{equation}
The function $\P{D}{2n}(2\lfl \eps n\rfl,2\lfl \delta\sqrt{n}\rfl)$ is the probability that 
the path passes through the point $(2\lfl \eps n \rfl,2\lfl \delta \sqrt{n} \rfl)$.  In order to
determine the probability density in the limit as $n\to\infty$, notice that the 
area element $\Delta A = \Delta m\, \Delta h$ has scaling with $n$ determined by 
$\Delta m = \lfl \eps n \rfl - \lfl \eps (n\minus 1)\rfl \propto \eps$
and $\Delta h = \lfl \delta\sqrt{n} \rfl - \lfl \delta \sqrt{n\minus 1} \rfl \propto
\delta/ \sqrt{n}$.  That is, the area element $\Delta A$ scales proportional to
$1/\sqrt{n}$.  By multiplying equation \Ref{13} by $\sqrt{n}$ and then taking
$n\to\infty$, the probability density
\begin{equation}
\Pr{D}(\eps,\delta) = \frac{4\,\delta^2\,e^{-\delta^2/\eps(1\minus\eps)}}{
\sqrt{\pi\, \eps^{3}(1\minus\eps)^{3}}}
\label{14}
\end{equation}
is obtained (see equation \Ref{4}).  This density induces a probability
measure $\rho(\eps,\delta)$ on $\RealN^2$ so that
$d\rho(\eps,\delta) = \Pr{D}(\eps,\delta)\,d\lambda$ where
$\lambda$ is plane measure.  The properties of $\Pr{D}(\eps,\delta)$ were already
discussed in section \ref{s1}.

An alternative (to the above) approach is to check the normalisation of equation
\Ref{13}.  Integrating this equation over $\delta$ gives
\begin{equation}
\int_0^\infty \P{D}{2n}(2\lfl \eps n\rfl,2\lfl \delta\sqrt{n}\rfl) \,d\delta = 1/\sqrt{n} + O(1/n)
\end{equation}
showing that the normalisation of equation \Ref{9} is lost when the 
subsitutions $m=\lfl \eps n\rfl$ and $h=\lfl\delta\sqrt{n}\rfl$ are made, followed
by an asymptotic expansion in $n$ (and the limit $n\to\infty$ is taken).
In particular, as pointed out above, this is due to the scaling of an area element
when the summation is approximated by an integral.  Thus, as $n\to\infty$,
equation \Ref{13} must be multiplied by $\sqrt{n}$ to recover the correct
normalisation of the probability measure in equation \Ref{14}. 

\subsection{Dyck meanders}
\label{s2.2}

\begin{figure}[h!]
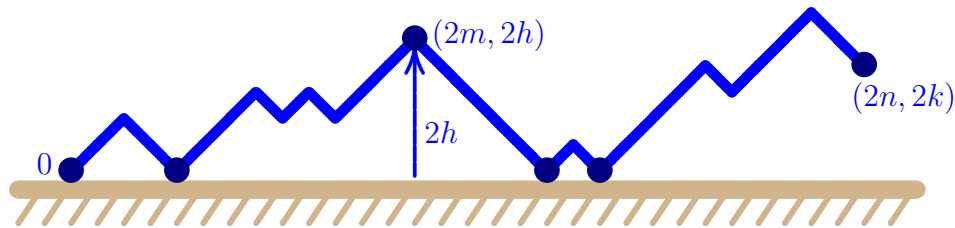

\beginpicture
\setcoordinatesystem units <1pt,1pt>
\setplotarea x from -70 to 250, y from -20 to 60

\setplotsymbol ({$\cdot$})
\arrow <10pt> [.2,.67] from 130 -2 to 130 47
\put {$0$} at -10 3
\put {$2h$} at 140 15
\put {$(2m,2h)$} at 158 52
\put {$(2n,2k)$} at 315 28

\color{Tan}
\setplotsymbol ({\Large$\bullet$})
\plot -20 -7 320 -7 /
\setplotsymbol ({\LARGE$\cdot$})
\plot -20 -20 -10 -5 / \plot -10 -20 0 -5 / \plot 0 -20 10 -5 / \plot 10 -20 20 -5 /
\plot 20 -20 30 -5 / \plot 30 -20 40 -5 / \plot 40 -20 50 -5 / \plot 50 -20 60 -5 /
\plot 60 -20 70 -5 / \plot 70 -20 80 -5 / \plot 80 -20 90 -5 / \plot 90 -20 100 -5 /
\plot 100 -20 110 -5 / \plot 110 -20 120 -5 / \plot 120 -20 130 -5 / \plot 130 -20 140 -5 /
\plot 140 -20 150 -5 / \plot 150 -20 160 -5 /

\plot 160 -20 170 -5 / \plot 170 -20 180 -5 / \plot 180 -20 190 -5 / \plot 190 -20 200 -5 /
\plot 200 -20 210 -5 / \plot 210 -20 220 -5 / \plot 220 -20 230 -5 / \plot 230 -20 240 -5 /
\plot 240 -20 250 -5 / \plot 250 -20 260 -5 / \plot 260 -20 270 -5 / \plot 270 -20 280 -5 /
\plot 280 -20 290 -5 / \plot 290 -20 300 -5 / \plot 300 -20 310 -5 / \plot 310 -20 320 -5 /

\color{Blue}
\setplotsymbol ({\footnotesize$\bullet$})
\plot 0 0  10 10  20 20  30 10 40 0 50 10 60 20 70 30 80 20
90 30 100 20 110 30 120 40 130 50 /
\plot 130 50 140 40 150 30 160 20 170 10 180 0 190 10 200 0 210 10
220 20 230 30 240 40 250 30 260 40 270 50 280 60 290 50 300 40 / 
\color{NavyBlue}
\multiput {\huge$\bullet$} at 0 0 40 0 130 50 180 0 200 0 300 40  /

\color{Black}
\normalcolor
\endpicture
\caption{A meander of length $2n$ passing through the vertex $(2m,2h)$.
The path is partitioned at this vertex into a meander, and then a directed
path from a vertex of height $2h$ to its endpoint at height $2k$.}
\label{f6}
\end{figure}

Generalising to Dyck meanders gives the model in figure \ref{f6} which
shows a meander ending in a vertex at height $2k$ having passed through the
vertex with coordinates $(2m,2h)$.   The case with an endpoint fixed at height
$2\lfl\omega \sqrt{n}\rfl$ is considered first, before the model with a free endpoint
is examined. An approach similar to that of the last section is followed, but
the extra degree of freedom in the model poses some additional challenges.

The number of meanders of length $2n$ from the origin to the 
site $(2n,2h)$ is given in equation \Ref{1}, while the number of directed
paths from a site at height $2h_0$ to a site at height $2h$, is given by
\begin{equation}
B(2n,2h_0,2h) = \Bi{2n}{n\plus h\minus h_0} - \Bi{2n}{n\plus h\plus h_0+1}.
\label{16}
\end{equation}
The probability that a path from the origin passes through the site $(2m,2h)$, 
conditioned to end in in the site $(2n,2k)$, is given by
\begin{equation}
\P{B}{2n}(2m,2h,2k) = \frac{D(2m,2h).B(2n\minus 2m,2h,2k)}{D(2n,2k)}.
\label{17}
\end{equation}
Putting $m=\lfl \eps n\rfl$, $h=\lfl \delta n \rfl$, and $k=\lfl \omega n \rfl$,
note that the asymptotics for $D(2\lfl \eps n \rfl,2\lfl \delta \sqrt{n}\rfl)$
were already calculated in equation \Ref{12}.  This also gives by substitution
an asymptotic formula for $D(2n,2\lfl \omega \sqrt{n} \rfl)$.  It remains to determine
an asymptotic formula for $B(2n\minus 2\lfl \eps n \rfl,2\lfl \delta \sqrt{n} \rfl,
2\lfl \omega \sqrt{n} \rfl)$.

More generally, consider $B(2\lfl \sigma n\rfl,2\lfl \delta\sqrt{n}\rfl,
2\lfl \omega \sqrt{n} \rfl)$ for $0< \sigma \leq 1$, and $\delta>0$
and $\omega>0$ (where $\sigma=1\minus\epsilon$).
Proceed by considering the first binomial coefficient in equation \Ref{16}.
Using Stirling's approximation to approximate the binomial coefficient,
taking logarithms, simplifying and taking an asymptotic expansion,
then exponentiating and simplifying gives 
(see equation \Ref{69} in Appendix A)
\begin{equation}
\hspace{-1.5cm}
\Bi{2\lfl\sigma n\rfl}{\lfl \sigma n \rfl +\lfl \omega \sqrt{n} \rfl - \lfl \delta \sqrt{n} \rfl}
=
\frac{e^{-(4(2\sigma n-1)(\delta-\omega)^2+\sigma)/(8\sigma^2 n)}}{
\sqrt{\pi\sigma n}} \,4^{\sigma n}
\, \LB 1 + O(\sigma^{-2}n^{-2}) \RB .
\label{18}
\end{equation}
Similarly, a slightly more complicated expression is obtained for the second
binomial coefficient in equation \Ref{16}:
\begin{eqnarray}
&\hspace{-2cm}
 \Bi{2\lfl\sigma n\rfl}{\lfl \sigma n \rfl +\lfl \omega \sqrt{n} \rfl + \lfl \delta \sqrt{n} \rfl+1} \cr
& =
\frac{e^{-(4n(2\sigma n-1)(\delta+\omega)^2+\sigma
+8(2\sigma n-1)(\delta+\omega)/\sqrt{n})/(8\sigma^2 n)}}{\sqrt{\pi\sigma n}}
\,4^{\sigma n} 
\, \LB 1 + O(\sigma^{-2}n^{-2}) \RB .
\label{19}
\end{eqnarray}
Subtracting equation \Ref{19} from equation \Ref{18} will give the
appropriate asymptotic approximation for $B(2\lfl \sigma n\rfl,2\lfl \alpha\sqrt{n}\rfl,
2\lfl \omega \sqrt{n} \rfl)$.

\begin{figure}[t]
\begin{center}
\includegraphics[width=0.675\textwidth]{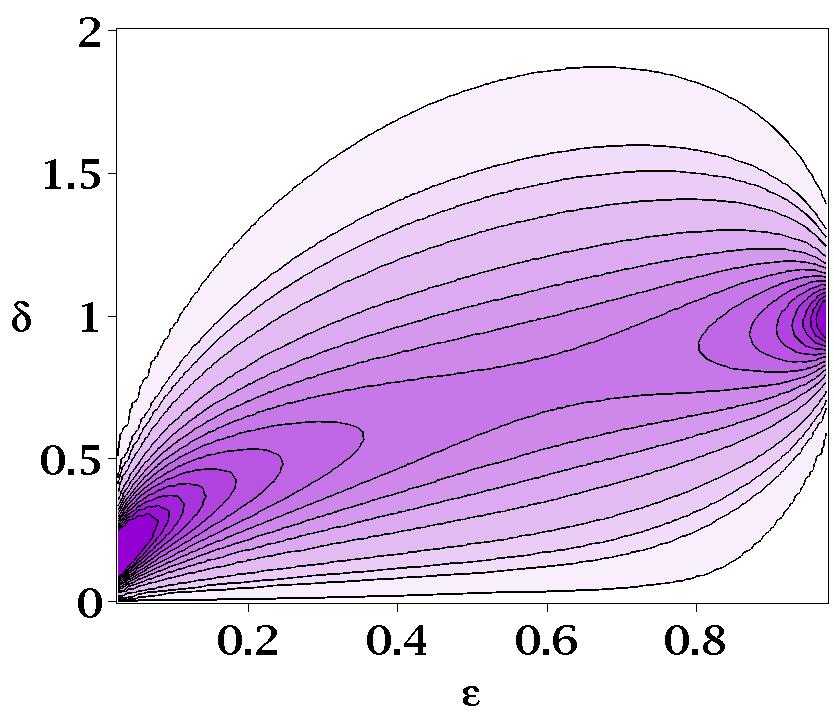}
\end{center}
\caption{The probability density $\Pr{B}(\eps,\delta,\omega)$ of a meander
passing through points in the $\eps\delta$-plane with $\omega=1$.}
\label{f7}
\end{figure}

Substituting the approximations in equations \Ref{18} and \Ref{19} in
equation \Ref{17} and using equation \Ref{12} 
give an approximation to $P_{2n}(2\lfl \eps n\rfl,2\lfl \delta n\rfl,
2\lfl \omega n \rfl)$.  Simplifying and taking the asymptotic expansion for large
$n$ and keeping only the leading order term gives
\begin{equation}
\hspace{-2.5cm}
\P{B}{2n}(2\lfl \eps n\rfl,2\lfl \delta \sqrt{n}\rfl, 2\lfl \omega\sqrt{n}\rfl)
= \frac{\delta\,(1-e^{-4\delta\omega/(1-\eps)})\,e^{-(\eps\omega-\delta)^2/\eps(1-\eps)}}{
\omega\sqrt{\pi n \eps^3(1-\eps)}}\,(1+O(1/\sqrt{n})).
\end{equation}
Normalising again by multiplying with $\sqrt{n}$, and then taking $n\to\infty$,
gives the probability density
\begin{equation}
\Pr{B} (\eps,\delta,\omega) 
= \frac{\delta\,(1-e^{-4\delta\omega/(1-\eps)})\,e^{-(\eps\omega-\delta)^2/\eps(1-\eps)}}{
\omega\sqrt{\pi \eps^3(1-\eps)}} .
\label{21}
\end{equation}
Integrating for $\delta>0$ shows that this is normalised.  A contourplot with
$\omega=1$ is shown in figure \ref{f7}.

The mean path can be calculated and is given by
\begin{eqnarray*}
\overline{\delta}(\eps) &= \int_0^\infty \delta\,\Pr{B}(\eps,\delta,\omega)\,d\delta \\
&= \frac{2\,\eps\omega^2-\eps+1}{2\omega} 
\,\erf\LB \omega\,\sqrt{\eps /(1\minus\eps)} \RB
+\sqrt{\eps(1\minus\eps) / \pi}\,e^{-\eps\omega^2/(1\minus\eps)} .
\end{eqnarray*}

The modal path is obtained by taking the derivative of equation \Ref{21}
to obtain an implicit expression relating $\delta$ to $\epsilon$:
\begin{equation}
e^{-4\delta\omega/(1-\eps)}
\LB 2\,\delta\eps\omega+2\,\delta^2+\eps^2-\eps\RB
+ \LB 2\,\delta\eps\omega-2\,\delta^2-\eps^2+\eps\RB = 0 .
\end{equation}
This can be simplified to
\begin{equation}
\tanh\LB \frac{2\,\delta\omega}{1-\eps} \RB 
= \frac{2\,\delta\eps\omega}{2\,\delta^2+\eps^2-\eps} .
\end{equation}
If $\omega$ is small, then the approximation $\tanh x \approx x - x^3/3$
can be used to obtain, after an expansion in $\omega$, followed by
exponentiation,
\begin{equation}
\delta_M(\eps) \approx \sqrt{\eps(1\minus\eps)}\,e^{\eps\omega^2/3(1-\eps)}.
\end{equation}
This approximation is accurate for small $\eps$, but breaks down as
$\eps \to 1^-$.  For large $\omega$ approximate $\tanh(x)\approx 1$ to obtain
\begin{equation}
\delta_M(\eps) \approx \Sfrac{1}{2}
\LB \eps\omega + \sqrt{(\omega^2\minus 2)\eps^2+2\,\eps} \RB .
\end{equation}
This approximation is in particular accurate for $\eps$ approaching $1$.

\begin{figure}[t]
\begin{center}
\includegraphics[width=0.675\textwidth]{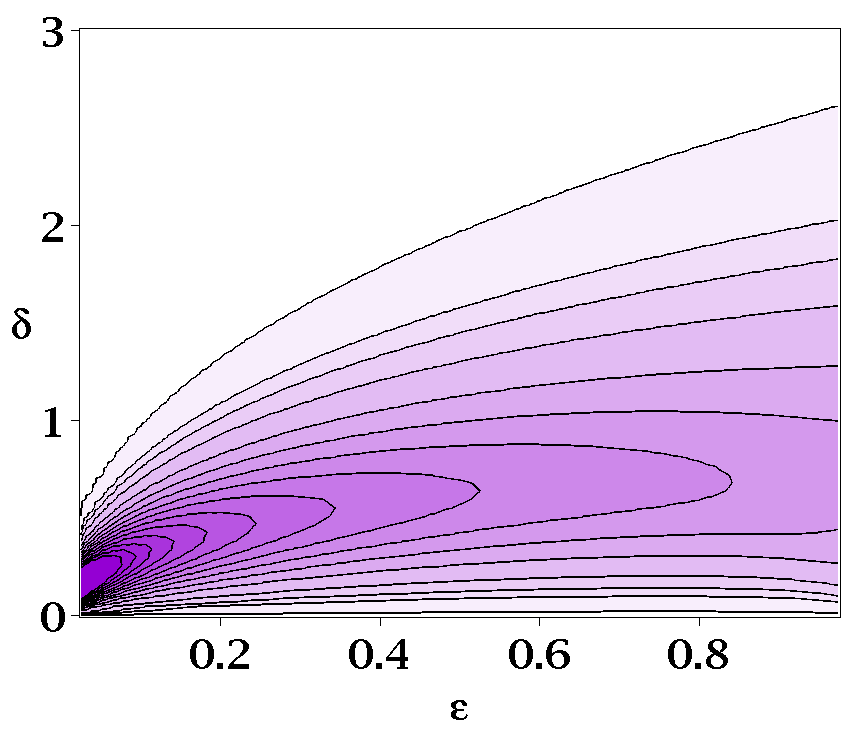}
\end{center}
\caption{The probability density $\Pr{F}(\eps,\delta,\omega)$ of a meander
with a free endpoint passing through points in the $\eps\delta$-plane.}
\label{f8}
\end{figure}

The density function can also be obtained for a model with a free endpoint.
Subtracting equation \Ref{19} from equation \Ref{18} gives an 
approximation for $B(\lfl \sigma n\rfl,\lfl\delta \sqrt{n}\rfl,\lfl\omega\sqrt{n}\rfl)$
with the same leading error terms as in Appendix A equation \Ref{69} where
$\lambda=\sigma$, $\delta=\delta$ and $\kappa=\omega$.
Since the error terms are constant, or decay to zero as $\omega\to\infty$,
$B(\lfl \sigma n\rfl,\lfl\delta \sqrt{n}\rfl,\lfl\omega\sqrt{n}\rfl)$
can be integrated over $\omega\in(0,\infty)$ for fixed $\delta>0$ to obtain 
an asymptotic approximation for the number of paths of length
$\lfl \sigma n \rfl$ starting at height $\lfl \delta \sqrt{n}\rfl$.

Proceed by first integrating the right hand side of equation \Ref{18} 
for $\omega\in(0,\infty)$.  This gives
\begin{equation*}
\hspace{-2.5cm}
\frac{4^{\sigma n}\sqrt{\sigma}\,e^{-1/8\sigma n}}{\sqrt{4n\sigma-2}}
\LB 1 + \erf\LB \frac{\delta\sqrt{2n\sigma-1}}{\sqrt{2n}\,\sigma} \RB \RB
= \frac{4^{\sigma n}}{\sqrt{4n}}
\LB 1 + \erf \LB \Sfrac{\delta}{\sqrt{\sigma}}\RB
+ O(1/\sqrt{n^3})  \RB .
\end{equation*}
Treating the right hand side of equation \Ref{19} the same way produces
\begin{eqnarray*}
& \hspace{-2.5cm}
\frac{4^{\sigma n}\sqrt{\sigma}\,e^{(7\sigma n - 4)/8\sigma^2 n^2}}{\sqrt{4n\sigma-2}}
\LB 1 - \erf\LB
\frac{2\sigma\sqrt{n}+2\sigma-8/\sqrt{n}-1/n}{\sigma\sqrt{4n\sigma-2}} \RB \RB \cr
&\hspace{1cm} = \frac{4^{\sigma n}}{\sqrt{4n}} \LB
1 - \erf\LB
\Sfrac{\delta}{\sqrt{\sigma}} \RB - \Sfrac{2\,e^{-\delta^2/\sigma}}{\sqrt{\pi\sigma n}}
+ O(1/\sqrt{n^3}) \RB .
\end{eqnarray*}
Substract the last equation from the penultimate to obtain 
\begin{eqnarray}
 F(\lfl \sigma n \rfl,\lfl\delta \sqrt{n} \rfl)
& = \int_0^\infty B(\lfl \sigma n \rfl,\lfl\delta \sqrt{n} \rfl,
\lfl \omega\sqrt{n}\rfl)\,d\omega \cr
& = \frac{4^{\sigma n}}{\sqrt{n}} \LB
\erf( \delta/\sqrt{\sigma})
+ \frac{e^{-\delta^2/\sigma}}{\sqrt{\pi \lambda n}} 
+ O(1/n) \RB .
\label{26}
\end{eqnarray}
The probability of a meander of length $2n$ from the origin, with a free endpoint,
passing through the point $(2m,2h)$ is
\begin{equation}
\P{F}{2n}(2m,2h) = \frac{D(2m,2h).F(2(n-m),2h)}{F(2n,0)} .
\end{equation}
Substituting $m=\lfl \eps n \rfl$ and $h=\lfl \delta\sqrt{n}\rfl$, and then
using equations \Ref{12} and \Ref{26} gives 
\begin{equation}
\P{F}{2n}(2\lfl \eps n\rfl, 2\lfl \delta \sqrt{n}\rfl)
= \frac{2\,\delta\,e^{-\delta^2/\eps}}{\sqrt{\eps^3}}\,
\erf(\delta/\sqrt{1-\eps}) + O(1/\sqrt{n}) .
\end{equation}
Taking $n\to\infty$ gives the probability density (see figure \ref{f8})
\begin{equation}
\Pr{F}(\eps,\delta) = \frac{2\,\delta\,e^{-\delta^2/\eps}}{\sqrt{\eps^3}}\,
\erf(\delta/\sqrt{1-\eps}).
\end{equation}
The normalisation of this probability density is also sound, since
\begin{equation}
\int_0^\infty \Pr{F}(\eps,\delta) \,d\delta = 1 . 
\end{equation}

The modal path is given by the solution $\delta_M(\eps)$ of
\begin{equation}
\erf\LB \frac{\delta}{\sqrt{1-\eps}} \RB
= \frac{2\,\delta\eps\,e^{-\delta^2/(1-\eps)}}{
(2\,\delta^2-\eps)\sqrt{\pi(1-\eps)}} .
\label{31}
\end{equation}
In the case that $\delta$ and $\eps$ are small, the modal path can be
approximated by
\begin{equation}
\delta_M(\eps) \approx \sqrt{\frac{3\,\eps}{2\,\eps+3}} 
= \sqrt{\eps} - \sqrt{\eps^3}/3 + \sqrt{\eps^5}/6 - \cdots.
\end{equation}
In the event that $\eps\to 1^-$, the path approaches $\delta(1)=1/\sqrt{2}$.
To see this, write equation \Ref{31} in the following form
\begin{equation}
\frac{e^{-\delta^2/(1-\eps)}}{\sqrt{\pi(1-\eps)}\; \erf(\delta/\sqrt{1-\eps})}
= \frac{2\,\delta^2-\eps}{2\,\delta\eps} .
\end{equation}
Take the left limit as $\eps\to 1^-$ shows that the left hand side approaches
zero, so that one must have $2\,\delta^2-1=0$ showing that 
$\delta_M(1^-)=1/\sqrt{2}$.

\begin{figure}[t!]
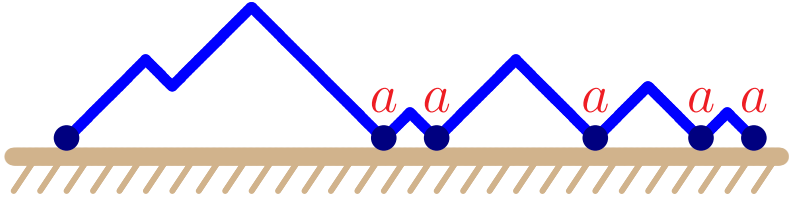

\beginpicture
\setcoordinatesystem units <1pt,1pt>
\setplotarea x from -80 to 250, y from -20 to 60

\color{Tan}
\setplotsymbol ({\Large$\bullet$})
\plot -20 -7 270 -7 /
\setplotsymbol ({\LARGE$\cdot$})
\plot -20 -20 -10 -5 / \plot -10 -20 0 -5 / \plot 0 -20 10 -5 / \plot 10 -20 20 -5 /
\plot 20 -20 30 -5 / \plot 30 -20 40 -5 / \plot 40 -20 50 -5 / \plot 50 -20 60 -5 /
\plot 60 -20 70 -5 / \plot 70 -20 80 -5 / \plot 80 -20 90 -5 / \plot 90 -20 100 -5 /
\plot 100 -20 110 -5 / \plot 110 -20 120 -5 / \plot 120 -20 130 -5 / \plot 130 -20 140 -5 /
\plot 140 -20 150 -5 / \plot 150 -20 160 -5 / \plot 160 -20 170 -5 / \plot 170 -20 180 -5 /
\plot 180 -20 190 -5 / \plot 190 -20 200 -5 / \plot 200 -20 210 -5 / \plot 210 -20 220 -5 /
\plot 220 -20 230 -5 / \plot 230 -20 240 -5 / \plot 240 -20 250 -5 / \plot 250 -20 260 -5 /
\plot 260 -20 270 -5 /

\color{Blue}
\setplotsymbol ({\footnotesize$\bullet$})
\plot 0 0  10 10  20 20  30 30 40 20 50 30 60 40 70 50 80 40
90 30 100 20 110 10 120 0 130 10 140 0 150 10 160 20 170 30
180 20 190 10 200 0 210 10 220 20 230 10 240 0 250 10 260 0   /
\color{NavyBlue}
\multiput {\huge$\bullet$} at 0 0 120 0 140 0 200 0 240 0 260 0  /
\color{Red}
\multiput {\LARGE$a$} at 120 15 140 15 200 15 240 15 260 15 /

\color{Black}
\normalcolor
\endpicture
\caption{A Dyck path model of an adsorbing linear polymer grafted on a surface.   
Each return (\textit{visit}) of the Dyck path the hard wall is weighted by an
activity $a$.  If $a$ is large then conformations of the path with a large number
of visits dominate the partition function, and the path is in an adsorbed state.}
\label{f9}
\end{figure}

\section{Adsorbing Dyck paths}
\label{s3}

\begin{figure}[h!]
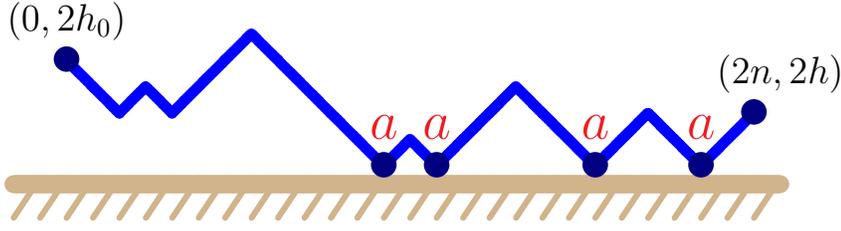

\beginpicture
\setcoordinatesystem units <1pt,1pt>
\setplotarea x from -80 to 250, y from -20 to 60

\color{Tan}
\setplotsymbol ({\Large$\bullet$})
\plot -20 -7 270 -7 /
\setplotsymbol ({\LARGE$\cdot$})
\plot -20 -20 -10 -5 / \plot -10 -20 0 -5 / \plot 0 -20 10 -5 / \plot 10 -20 20 -5 /
\plot 20 -20 30 -5 / \plot 30 -20 40 -5 / \plot 40 -20 50 -5 / \plot 50 -20 60 -5 /
\plot 60 -20 70 -5 / \plot 70 -20 80 -5 / \plot 80 -20 90 -5 / \plot 90 -20 100 -5 /
\plot 100 -20 110 -5 / \plot 110 -20 120 -5 / \plot 120 -20 130 -5 / \plot 130 -20 140 -5 /
\plot 140 -20 150 -5 / \plot 150 -20 160 -5 / \plot 160 -20 170 -5 / \plot 170 -20 180 -5 /
\plot 180 -20 190 -5 / \plot 190 -20 200 -5 / \plot 200 -20 210 -5 / \plot 210 -20 220 -5 /
\plot 220 -20 230 -5 / \plot 230 -20 240 -5 / \plot 240 -20 250 -5 / \plot 250 -20 260 -5 /
\plot 260 -20 270 -5 /

\color{Blue}
\setplotsymbol ({\footnotesize$\bullet$})
\plot 0 40  10 30  20 20  30 30 40 20 50 30 60 40 70 50 80 40
90 30 100 20 110 10 120 0 130 10 140 0 150 10 160 20 170 30
180 20 190 10 200 0 210 10 220 20 230 10 240 0 250 10 260 20   /
\color{NavyBlue}
\multiput {\huge$\bullet$} at 0 40 120 0 140 0 200 0 240 0 260 20  /
\color{Red}
\multiput {\LARGE$a$} at 120 15 140 15 200 15 240 15 /

\color{Black}
\put {\large$(0,2h_0)$} at 0 55   \put {\large$(2n,2h)$} at 270 35
\normalcolor
\endpicture
\caption{An adsorbing directed path model of an adsorbing linear polymer.
The path starts in a site at height $2h_0$, and makes $2n$ steps to its
endpoint at height $2h$. Each return of the path to the hard wall is a 
visit and is weighted by the activity $a$.}
\label{f10}
\end{figure}

Adsorbing Dyck paths (see figure \ref{f9}) is a model of an adsorbing linear polymer \cite{BEO98,BEO99,BGW91,BGW92,RJvR04} with monomers which may stick to
the adsorbing surface.  The generating function of adsorbing
Dyck paths is given by
\begin{equation}
D(a,t) = \frac{2}{2-a\,(1-\sqrt{1\minus 4t^2})},
\label{34}
\end{equation}
where $t$ is the generating variable for steps in the path, and $a$ is the
generating variable of returns (visits) of the path to the adsorbing (hard) wall.
In order to analyse this model, consider directed paths of length $2n$ from a 
site $(0,2h_0)$ to a site $(2n,2h)$ as shown in figure \ref{f10}.  The
partition function is given by (see equation (5.33) in reference \cite{JvR15})
\begin{eqnarray}
d_n(h_0,h) &= \Bi{2n}{n\plus h\minus h_0} - \Bi{2n}{n\plus h\plus h_0}
\label{35} \\
&\quad
+ a \sum_{\ell=0}^{n\minus h\minus h_0}
\LB \Bi{2n}{n\plus h\plus h_0\plus \ell} 
   - \Bi{2n}{n\plus h\plus h_0\plus \ell \plus 1} \RB (a-1)^\ell . \nonumber
\end{eqnarray}
The difference between the first two binomial coefficients counts paths
having no intersection with the adsorbing wall, while the summation over
$\ell$ counts paths intersecting the adsorbing wall at least once.

The generating function in equation \Ref{34} shows a critical adsorption
point at the critical activity $a_c=2$, and this is the \textit{special point}.  
For $a>2$ the model is in an adsorbed state, with paths intersecting the adsorbing 
wall dominating the partition function.  If $a<2$, then the first two binomial 
coefficients dominate equation \Ref{35}, and the path is desorbed.

\subsection{The density of adsorbed directed paths (for $a>2$)}
\label{s3.1}

Since we will ultimately only be interested in paths starting at the
origin (so that $h_0=0$), the first two binomial coefficients can
be ignored in equation \Ref{35}.  

Put $A=a\minus 1$ and denote the summand in equation \Ref{35} by
\begin{equation}
S(\ell) = 
\frac{1}{n\plus h_0\plus h\plus \ell\plus 1}
\Bi{2n}{n\plus h_0\plus h\plus \ell}\,A^\ell .
\label{36}
\end{equation}
Substitute $n=\lfl \sigma m^2\rfl$ and $h_0\plus h=\lfl \alpha m\rfl$.  Since 
the summation over $\ell$ in equation \Ref{35} has a range of $O(n)$,
put $\ell = \lfl \tau m^2\rfl$.  Use the approximation in equation \Ref{10}
to approximate the factorials, and after taking logarithms and expanding
one obtains
\begin{eqnarray*}
&\hspace{-1cm} (1/2+2\,\sigma m^2)(\log(\sigma)+\log(m^2))
-(1/2)\log(m(\sigma\sqrt{n}+\tau m+\alpha)) \\
&\hspace{-1cm}+(-\tau n-\alpha m)\log(\sigma m^2+\tau m^2+\alpha m)
-(1/2)\log (m(\sigma m-\tau m-\alpha))\\
&\hspace{-1cm}+(-\sigma m^2+\tau m^2+\alpha m) \log(\sigma m^2-\tau m^2-\alpha m)
-\log (\sigma m^2+\tau m^2+\alpha m+1)\\
&\hspace{-1cm}+\tau m^2\log A-\sigma m^2\log (\sigma m^2+\tau m^2+\alpha m)/4) 
- (1/2)\log\pi .
\end{eqnarray*}
Expanding asymptotically in $m$, exponentiating the result, and simplifying give
\begin{equation}
\frac{A^{\tau m^2} 4^{\sigma m^2} \sigma^{2\sigma m^2 + 1/2}}{
\sqrt{\pi}\,(\sigma+\tau)\,(\sigma^2-\tau^2)^{\sigma m^2}\,m^3}
\LB \frac{\sigma-\tau}{\sigma+\tau} \RB^{\tau m^2+\alpha m + 1/2}
\; e^{-\sigma\alpha^2/(\sigma^2-\tau^2)} .
\label{37}
\end{equation}
This must be integrated for $\tau\in[0,\sigma]$ to approximate the
summation over $\ell$ in equation \Ref{35}.  The resulting integral is
best estimated using a saddle point approximation \cite{A67}, namely
\begin{equation}
\int_0^\infty e^{\lambda\, F(t)}\,G(t)\,dt \approx
\frac{\sqrt{2\pi}\; e^{\lambda\, F(t_0)}\,G(t_0) }{\sqrt{-\lambda\, F^{\pr\pr}(t_0)}},
\end{equation}
where $t_0$ is the \textit{saddle point} given by the solution of $F^\pr(t)=0$.
Since we are interested in the asymptotic regime, and will be taking $m\to\infty$,
identify the function $\lambda\, F(t)$ by collecting all factors containing $m^2$:
\begin{equation}
e^{\lambda\,F(m)} 
= \frac{\sigma^{2\sigma m^2}A^{\tau m^2}4^{\sigma m^2}}{
(\sigma+\tau)^{m^2(\sigma+\tau)}\,(\sigma-\tau)^{m^2(\sigma-\tau)}} .
\end{equation}
Taking logarithms and differentiating, and then solving for the stationary 
value of $\tau$ gives the location of the saddle point:
\begin{equation}
\tau_s = \LB \frac{A-1}{A+1} \RB \sigma = \LB \frac{a-2}{a} \RB \sigma .
\label{40}
\end{equation}
Since one must have $0<\tau_s<\sigma$, this result shows that 
the approximation is only valid if $a>2$, or in fact if
$A>1$.  Since $a=2$ is the special point (the critical adsorption point), our
approximation will only be valid in the adsorbed phase of the model.
Substituting $\tau=\tau_s$ gives the function $G(\tau_s)$ in the saddle
point approximation:
\begin{equation}
G(\tau_s) = \frac{(A+1)^2\,A^{-\alpha m - 3/2}}{4 \sqrt{\pi}\,\sigma^{3/2}\,m^3}
\, e^{-\alpha^2(A+1)^2/4\sigma A} .
\end{equation}
Calculating $\lambda F(\tau_s)$ and using the saddle point approximation
then gives the following approximation of equation \Ref{35} when
$m^2 = n$ (notice that the original substitution was $n = \lfl \sigma m^2\rfl$)
and $h+h_0 = \alpha \sqrt{n}$:
\begin{equation}
f(\sigma,\alpha) = \LB \frac{(A+1)^2}{A} \RB^{\sigma n + 1/2}
\frac{e^{-\alpha^2(A+1)^2/\sigma A}}{2\sigma n^2\,A^{\alpha\sqrt{n}+1/2}} .
\end{equation}
The function $f(\sigma,\alpha)$ is a finite size approximation to the
partition function of directed paths of length $\lfl 2\sigma n\rfl$, adsorbing 
into a hard wall, and with the combined height of its endpoints 
$h_0+h = \lfl \alpha \sqrt{n} \rfl$.

\begin{figure}[t]
\includegraphics[width=0.33\textwidth]{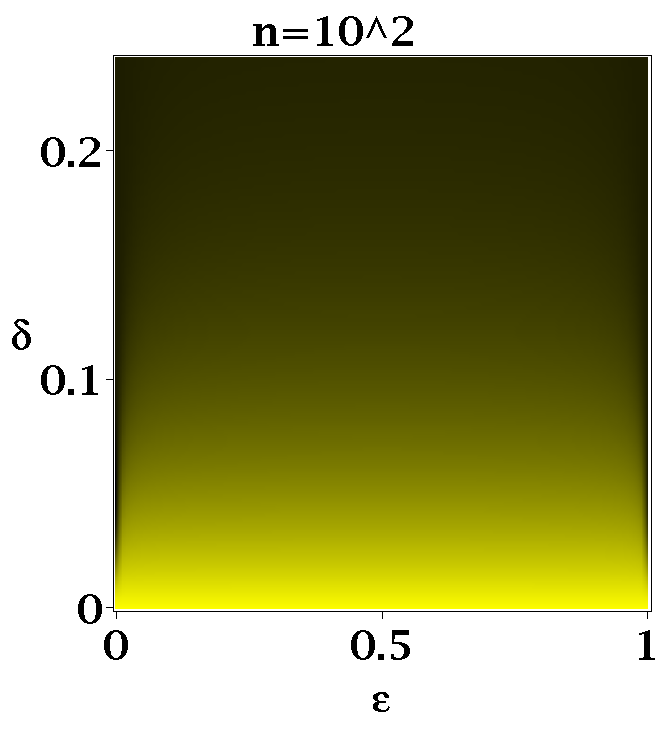}
\includegraphics[width=0.33\textwidth]{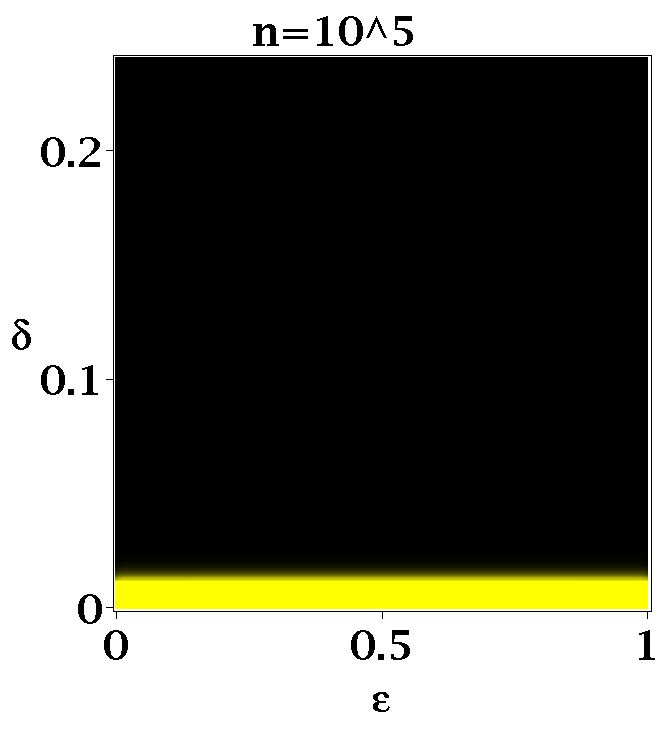}
\includegraphics[width=0.33\textwidth]{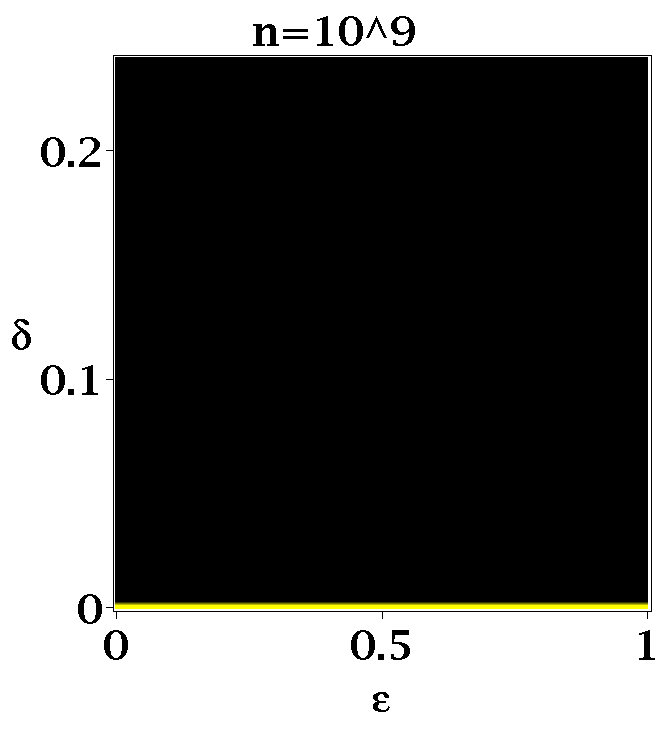}
\caption{The probability $\P{C}{2n}(2\lfl \eps n\rfl,2\lfl \delta\rfl)$ 
of an adsorbing Dyck path with activity $a=3$.}
\label{F7}
\end{figure}

The probability that the path passes through the point $(2\lfl \eps n\rfl,
2\lfl \delta \sqrt{n} \rfl)$ having started in the origin, and then
ending in the adsorbing wall, is
\begin{eqnarray}
\hspace{-2cm}
\P{C}{2n}(2\lfl \eps n\rfl,2\lfl \delta \sqrt{n} \rfl) 
& \approx \frac{f(\eps,\delta).f(1{-}\eps,\delta)}{f(1,0)} \nonumber \\
& = \LB\frac{(A+1)}{2n^2\,\eps(1-\eps)}\RB A^{-2\,\delta\sqrt{n}-1}
e^{-\delta^2(A+1)^2/4\eps(1\minus\eps)A} .
\end{eqnarray}
Due to the approximations used to derive this expression, it is not
normalised, and must be, so integrate $\delta\in[0,\infty)$.  This
produces an expression containing error functions, which is again
expanded asymptotically.  Substituting and then simplifying
gives the final approximation of the finite size probability:
\begin{equation}
\P{C}{2n}(2\lfl \eps n\rfl,2\lfl \delta \sqrt{n} \rfl)  
\simeq 2\sqrt{n}\,(\log A) \,A^{-2\,\delta\sqrt{n}}\,
e^{-\delta^2(A+1)^2/4\eps(1\minus\eps)A} .
\label{44}
\end{equation} 
As expected, this result breaks down when $a=2$ (or $A=1$).
In figure \ref{F7} three density plots are shown for 
$n=10^2$, $n=10^5$ and $n=10^9$.  Notice the narrow (vertical) range
of $\delta$.  The density concentrates towards $\delta=0$ with increasing
$n$, from a more diffused density when $n=10^2$, to a very narrow
bar along $\delta=0$ when $n=10^9$.  These plots are for $a=3$.

Consider next a closed rectangle $R = [\eps_1,\eps_2]\times [\delta_1,\delta_2]$
with $0 \leq \eps_1 < \eps_2\leq 1$ and $0 < \delta_1< \delta_2 <\infty$.
The probability that the path passes through $R$ is given by
\begin{equation}
\hbox{Prob}_n(R) = \int_R \P{C}{2n}(2\lfl \eps n\rfl,2\lfl \delta \sqrt{n} \rfl) \, d\lambda
+ \hbox{[finite size corrections]},
\end{equation} 
up to finite size corrections which decay as $n$ approaches infinity, and 
where $\lambda$ is plane measure.  Since $\delta_1>0$, it follows that
\begin{eqnarray}
\lim_{n\to\infty} \hbox{Prob}_n(R)
&= \lim_{n\to\infty} \int_R 
\P{C}{2n}(2\lfl \eps n\rfl,2\lfl \delta \sqrt{n} \rfl) \, d\lambda \nonumber \\ 
&\leq \lim_{n\to\infty} \P{C}{2n}(2\lfl \delta_1 \sqrt{n} \rfl,0)\times \hbox{Area}(R) = 0.
\label{48}
\end{eqnarray}
This result is true even in the limit that $\delta_2\to\infty$. In other words, the probability 
that the path passes through the infinite rectangle $[\eps_1,\eps_2]\times [\delta_1,\infty)$
is zero, for any $\delta_1>0$, and for $0\leq \eps_1 < \eps_2 \leq 1$.

If $\delta_1=0$, then consider the rectangle 
$R = [\eps_1,\eps_2]\times [0,\delta_2)$. Define the infinite rectangle
$R_\infty = [\eps_1,\eps_2]\times [0,\infty)$ and put
$R_2 = [\eps_1,\eps_2]\times [\delta_2,\infty) = R_\infty\setminus R$. 
Then it follows that
\begin{equation}
\lim_{n\to\infty} \hbox{Prob}_n(R_\infty) = \eps_2-\eps_1
\end{equation} 
since $\P{C}{2n}(2\lfl \eps n\rfl,2\lfl \delta \sqrt{n} \rfl)$ in equation \Ref{44} 
is normalised in the $n\to\infty$ limit.  However, by equation \Ref{48},
since $R_\infty = R\cup R_2$ and $R\cap R_2=\emptyset$,
\begin{equation}
\hspace{-2.5cm}
\lim_{n\to\infty} \hbox{Prob}_n(R) 
= \lim_{n\to\infty} \hbox{Prob}_n(R_\infty) 
   - \lim_{n\to\infty} \hbox{Prob}_n(R_2)
= \lim_{n\to\infty} \hbox{Prob}_n(R_\infty)
= \eps_2-\eps_1 .
\end{equation} 
Since this result is independent of $\delta_2>0$, take the limit
$\delta_2\to 0^+$, and define the set function $\xi(R)$ on rectangles
$R$ as follows:  If $R=[\eps_1,\eps_2]\times[c_1,c_2]$, then
\begin{equation}
\hspace{-2cm}
\xi(R) = \lim_{n\to\infty} \hbox{Prob}_n(R) =
\cases{
0, 
& \hbox{if $0\leq \eps_1<\eps_2\leq 1$ and $0 < c_1 < c_2$}; \\
\eps_2-\eps_1, 
& \hbox{if $0\leq \eps_1<\eps_2\leq 1$ and $0 = c_1 < c_2$}; \\
0, &\hbox{otherwise}.}
\end{equation}
Then $\xi$ is a set-function on rectangles in the plane $\RealN^2$ and 
an outer measure $\xi$ on sets $E\subseteq \RealN^2$ can be defined by
\begin{equation}
\hspace{-1cm}
\xi\, E = \inf \LC \sum_{j} \xi R_j
\vv \hbox{$E\subseteq \cup R_j$ and $\{R_j\}$ is a countable cover of $E$} \RC,
\end{equation}
where the infimum is taken over all countable covers of $E$ by rectangles.
Since $\xi\,\RealN^2 = 1$, $\xi$ induces a probability measure $\Pr{C}$ on 
the $\sigma$-algebra of plane measure sets, such that the limiting 
probability that a path passes through points in $E$ is given by
\begin{equation}
\Pr{C}(E) = \int_E d\xi = \int_E \frac{d\xi}{d\lambda}\,d\lambda 
= \int_E \Pr{C}(\eps,\delta)\,d\lambda 
\label{50}
\end{equation}
so that $d\xi = \Pr{C} (\eps,\delta)\,d\lambda$ and $\Pr{C} (\eps,\delta)$ is the Radon-Nikodym
derivative of $\xi$ with respect to plane measure $\lambda$ (and is the probability
density of the model in the limit $n\to\infty$).  Notice that $\Pr{C}(\eps,\delta) = 0$ 
everywhere in the $\eps\delta$-plane except on the line segment $\delta=0$ 
and $\eps\in[0,1]$ but that it is normalised such that for $\delta>0$, if
$R_\delta = [0,1]\times[0,\delta]$,
\begin{eqnarray}
\hspace{-2cm}
\int_{R_\delta}  \Pr{C}(\eps,\delta)\,d\lambda &= 
\lim_{\delta\to 0^+} \int_{R_\delta}  \Pr{C}(\eps,\delta)\,d\lambda \cr
&= 
\int_0^1\int_0^\infty \Pr{C}(\eps,\delta)\,d\delta\,d\eps = 
\int_{\RealN^2} \Pr{C}(\eps,\delta)\,d\lambda = 1.
\end{eqnarray}

\subsection{Paths at the special point}
\label{s3.2}

At the critical (special) point $a=2$ equation \Ref{35} telescopes to
\begin{equation}
d_n(h_0,h)\vv_{a=2} = \Bi{2n}{n+h-h_0} + \Bi{2n}{n+h+h_0} - 4n.
\end{equation}
Substitute $n=\lfl\sigma m^2\rfl$, $h_0=\lfl \alpha \sqrt{n} \rfl$ and
$h=\lfl\omega\sqrt{n} \rfl$.  In order to determine the asymptotic
behaviour at $a=2$, consider the binomial terms above independently.
As before, use equation \Ref{10} to approximate factorials,
take logarithms, simplify and expand the resulting expressions asymptotically.
Exponentiation the result gives
\begin{equation}
\Bi{2\lfl\sigma m^2\rfl}{\lfl\sigma m^2\rfl+\lfl\omega m \rfl-\lfl \alpha m \rfl}
\simeq 
\frac{4^{\sigma m^2}}{\sqrt{\pi\sigma}\,m}\;
e^{-(\alpha-\omega)^2(6m\sigma^2-3\sigma+(\alpha-\omega)^2)/6\sigma^3m^2}.
\end{equation}
Similarly,
\begin{equation}
\Bi{2\lfl\sigma m^2\rfl}{\lfl\sigma m^2\rfl+\lfl\omega m \rfl+\lfl \alpha m \rfl}
\simeq 
\frac{4^{\sigma m^2}}{\sqrt{\pi\sigma}\,m}\;
e^{-(\alpha+\omega)^2(6m\sigma^2-3\sigma+(\alpha+\omega)^2)/6\sigma^3m^2}.
\end{equation}
Combining these expressions, simplifying, taking logarithms, expanding 
asymptotically in $m$, replacing $m^2=n$, and then exponentiating gives to
leading order
\begin{equation}
\hspace{-2.5cm}
\frac{4^{\sigma n}e^{-(\alpha^2+\omega^2)/\sigma}}{\sqrt{\pi\sigma n}}
 \LB e^{2\alpha\omega/\sigma}
\plus e^{-2\alpha\omega/\sigma} \RB - 4\sigma n
= 
\frac{4^{\sigma n+1/2}e^{-(\alpha^2+\omega^2)/\sigma}}{\sqrt{\pi\sigma n}}\,
 \cosh(2\alpha\omega/\sigma)
- 4\sigma n.
\end{equation}
The last term $4\sigma n$ can be ignored in the asymptotic limit.  Thus, put
\begin{equation}
f(\sigma,\alpha,\omega) = 
\frac{4^{\sigma n+1/2}e^{-(\alpha^2+\omega^2)/\sigma}}{\sqrt{\pi\sigma n}}\,
 \cosh(2\alpha\omega/\sigma).
\end{equation}
Using the same arguments as in section \ref{s2.2}, the probability that the path passes 
through the point $(2\lfl\eps n\rfl,2\lfl\delta \sqrt{n}\rfl)$ is asymptotically 
approximated by
\begin{equation}
\P{S}{2n}(2\lfl\eps n\rfl,2\lfl\delta \sqrt{n}\rfl) = \frac{2\,e^{-\delta^2/\eps(1-\eps)}}{\sqrt{\pi\eps(1-\eps)}} .
\end{equation}
Notice that all $n$-dependence has dropped out to leading order, so that this is also the 
probability density of the model.  That is, taking $n\to\infty$ gives the density
\begin{equation}
\Pr{S} (\eps,\delta) = \frac{2\,e^{-\delta^2/\eps(1-\eps)}}{\sqrt{\pi\eps(1-\eps)}} .
\end{equation}

\begin{figure}[t]
\begin{center}
\includegraphics[width=0.675\textwidth]{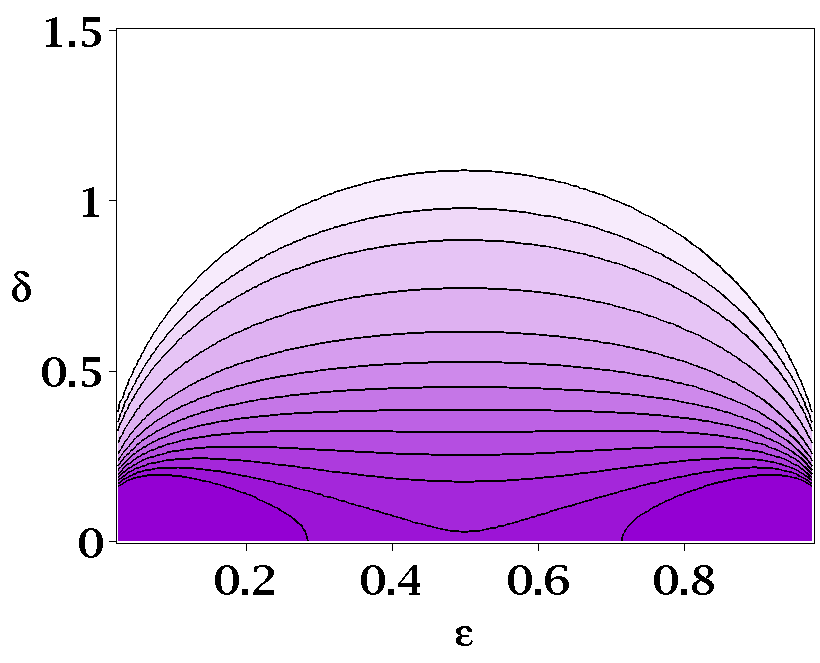}
\end{center}
\caption{The probability density $\Pr{S}(\eps,\delta)$ of an adsorbing
Dyck path at the special point.}
\label{F6}
\end{figure}

The mean path is given by
\begin{equation}
\overline{\delta}(\eps) = \int_0^\infty \delta\, \Pr{S} (\eps,\delta)\, d\delta
= \sqrt{\eps(1-\eps)/\pi} .
\end{equation}

The modal path is obtained by solving $\sfrac{d}{d\delta} \log \rho (\eps,\delta) 
= 2\,\delta / \eps(1\minus \eps) = 0$.  This shows 
\begin{equation}
\delta_M(\eps) = 0, \quad\hbox{for $\eps\in[0,1]$}.
\end{equation} 
This is expected, since at the critical point the path is adsorbed on the wall.  
The median path is obtained by solving for $\delta_m(\eps)$ in
\begin{equation}
\int_0^{\delta_m} \Pr{S} (\eps,\delta)\,d\delta
= \erf\LB \delta_m/\sqrt{\eps(1\minus\eps)} \RB = 1/2.
\end{equation}
A numerical solution gives
\begin{equation}
\delta_m(\eps) = \z0\,\sqrt{\eps(1\minus\eps)} \approx 0.4769 \,
\sqrt{\eps(1\minus\eps)}
\end{equation}
where $\z0$ is the real solution of $2\,\erf(z)=1$.

\subsection{The case that $a<2$.}
\label{s3.3}

In the event that $a<2$ the saddle point $\tau_2$ in equation \Ref{40} moves
to a negative value.  This shows, asymptotically, that $\tau=0$ maximizes
equation \Ref{37}, and therefore that the $\ell=0$ term in equation \Ref{36}
is dominant in the asymptotic regime.  Thus, putting $h_0=0$ in equation
\Ref{35}, gives the approximation
\begin{eqnarray*}
D(2n,2h) &= d_{2n}(0,2h)
\lesssim a \sum_{\ell=0}^\infty \frac{1}{n+h+\ell+1} \Bi{2n}{n+h+\ell}\,A^\ell \\
&\lesssim  \frac{a}{2-a}\;  \frac{1}{n+h+1} \Bi{2n}{n+h} ,
\end{eqnarray*}
since $A=a\minus 1$. Comparison to equation \Ref{1} and then using equation
\Ref{3} gives expressions similar to equation \Ref{9} for the probability
$\P{D}{2n}(2m,2h)$ passing through the point $(2m,2h)$.
Determining the asymptotics of this gives equation \Ref{14}.

\section{Discussion}
\label{s4}

In this paper the probability distributions of directed and adsorbing directed paths
were determined.  The results are exact, but were not proven with full mathematical
rigour (that will require the tracking of error terms in the asymptotic approximations).
We collect our results in table \ref{table-density} below.

\renewcommand{\arraystretch}{1.25}
\begin{table}[h]
\caption{Probability densities of adsorbing directed paths}
\centering
\scalebox{1.0}{        
\begin{tabular}{l | @{}*{1}{l}}
\br    
Model &\qquad $\mathbb{P}(\eps,\delta,\ldots)$ \cr 
\mr
Dyck paths  \quad &\quad  $\frac{4\delta^2}{\sqrt{\pi\,\eps^3(1\minus\eps)^3}}\, e^{-\delta^2/\eps(1\minus\eps)}$  \cr
Meanders \quad &\quad  $\frac{\delta}{\omega\sqrt{\pi\,\eps^3(1\minus\eps)}}
\, (1\minus e^{-4\delta\omega/(1\minus\eps)}) 
\, e^{-(\eps\omega-\delta)^2/\eps(1\minus\eps)}$ \cr
Meanders (free endpoint) \quad &\quad $\frac{2\delta}{\sqrt{\eps^3}}
\, e^{-\delta^2/\eps}\,\erf(\delta/\sqrt{1\minus\eps})$ \cr
Dyck paths (adsorbed phase) \quad&\quad \hbox{via equation \Ref{50}} \cr
Dyck paths (special point) \quad&\quad $\frac{2}{\sqrt{\pi \eps(1\minus\eps)}}
\, e^{-\delta^2/\eps(1\minus\eps)}$ \cr
\br
\end{tabular}
}
\label{table-density}   
\end{table}

In the desorbed phases these models have distributions peaking away from the hard
wall, consistent with grafted directed polymers tending to drift away from the hard wall
forming areas of lower density close to the hard wall, and higher density further away.
This is seen in figure \ref{f5} for Dyck paths, and also in figures \ref{f7} and \ref{f8},
for models of meanders.  In the context of figure \ref{f1}, this shows that particles
close to the hard wall has to overcome entropic repulsion to
migrate away from the hard wall, and this is the mechanism underlying the absorption
and stabilisation of molecules by a system of linear polymers grafted to a hard wall,
for example in drug eluding stents or nanoparticle-polymer systems.  Observe that in
the adsorbed phase the polymer is fully adsorbed, and this mechanism is not operative.
This is also the case at the special point, where the probability density decreases 
monotonically with distance from the hard wall.

\section*{Appendix A}

In this appendix we give a summary of the derivation of equations \Ref{18} and \Ref{19}.

The asymptotic approximation of the factorial can be determined using a symbolic 
computations program \cite{maple17}:
\begin{equation}
n! = \sqrt{2\pi n}\, n^n e^{-n} \LB 1 + \Sfrac{1}{12\,n} 
+ \Sfrac{\xi}{n^2} \RB,
\quad\hbox{where $\xi = c_0 + o(1)$.}
\end{equation}
Here, $\xi$ is a function of $n$ and it may be verified that $\xi=c_0+O(1/n)$
where $c_0$ is a non-zero constant.

Noting that $\sqrt{1+1/6n} = 1 + 1/12n + O(1/n^2)$, this may be written 
in the form
\begin{equation}
n! = \sqrt{\pi(2n + 1/3)}\, n^n e^{-n}\,\LB 1 +\Sfrac{\zeta}{n^2}\RB,
\quad\hbox{where $\zeta = c_1 + o(1)$}
\label{64}
\end{equation}
where $c_1$ is again a non-zero constant and $\zeta=c_1+O(1/n)$.
This gives equation \Ref{10}.

Next,  consider the left hand side of equation \Ref{18}.  This is of the generic
form
$$A = \scalebox{0.9}{
$\displaystyle{
\Bi{2 \lfl \lambda n\rfl}{
       \lfl\lambda n\rfl + \lfl \delta\sqrt{n}\rfl + \lfl \kappa\sqrt{n}\rfl + c}
=
\frac{(2\lfl \lambda n\rfl) !}{
   (\lfl \lambda n\rfl + \lfl \delta\sqrt{n}\rfl + \lfl \kappa\sqrt{n}\rfl + c)!\,
   (\lfl \lambda n\rfl - \lfl \delta\sqrt{n}\rfl - \lfl \kappa\sqrt{n}\rfl - c)!}
}$}$$
To obtain equation \Ref{18} put $c=0$ and change the sign of 
$\lfl \kappa\sqrt{n}\rfl$, and for equation \Ref{19} put $c=1$.

Approximate the factorials using equation \Ref{64}.  Making the following
substitutions
\begin{eqnarray*}
a_1 &= \pi (4\lambda n+1/3), \cr
a_2 &= (\delta+\kappa) \sqrt{n}+\lambda n+c, \cr
a_3 &= -(\delta+\kappa)\sqrt{n}-\lambda n-c, \cr 
a_4 &= \pi(2(\delta+\kappa) \sqrt{n}+2\lambda n+2c+1/3), \cr
a_5 &= -(\delta+\kappa) \sqrt{n}+\lambda n-c, \cr
a_6 &= (\delta+\kappa) \sqrt{n} - \lambda n+c, \cr
a_7 &= \pi (-2(\delta+\kappa) \sqrt{n}+2 \lambda n-2c+1/3), \cr
a_8 &= 2\lambda n,
\end{eqnarray*}
gives
\begin{equation}
A = \frac{\sqrt{a_1}\, a_8^{a_8}\, e^{-2\lambda n}\, (1+ \zeta/(4\lambda^2 n^2))
}{\sqrt{a_4}\, a_2^{a_2}\, e^{a_3} \,(1+\zeta/a_2^2)
   \; \sqrt{a_7}\, a_5^{a_5}\, e^{a_6} \, (1+\zeta/a_5^2) }
\end{equation}
Take the logarithm of $A$, and expand asymptotically in $n$.
The terms depending on $\zeta$ give the error, and this is, asymptotically in $n$,
\begin{equation}
\hspace{-2cm}
e^{\hbox{Error}} 
= \frac{(1+ \zeta/(4\lambda^2 n^2))}{(1+\zeta/a_2^2) \, (1+\zeta/a_5^2) }
= 1 - \frac{7\zeta}{4\lambda^2 n^2} - \Sfrac{6\zeta(\delta+\kappa)^2}{\lambda^4n^3}+ O\LB \lambda^{-4} n^{-7/2} \RB . 
\end{equation}
This approaches $1$ as $n\to\infty$ for any fixed $\lambda>0$.  Taking the logarithm
of $A$ and expanding asymptotically in $n$ gives the result
\begin{eqnarray*}
\log A 
&= \lambda n \log 4-(\log \pi)/2
-2\delta\kappa/\lambda-\kappa^2/\lambda-(\log \lambda)/2-\delta^2/\lambda \cr
& -(\log n)/2-2(\delta+\kappa) c/(\lambda \sqrt{n})
   +(4\delta^2+8\delta\kappa+4\kappa^2-\lambda)/(8\lambda^2 n) \cr
& +(\delta+\kappa)c/(\lambda^2 n^{3/2}) + \hbox{Error}
\end{eqnarray*}
Exponentiating and simplifying then produces
\begin{equation}
A = \frac{
e^{-
(8\lambda(\delta+\kappa)^2n^{3/2}+16c\lambda(\delta+\kappa)n
    -(4(\delta+\kappa)^2-\lambda)\sqrt{n}-8c (\delta +\kappa))
/(8\lambda^2 n^{3/2})
}
e^{\hbox{\footnotesize Error}}
}{
\sqrt{\pi\lambda n}
}
\,4^{\lambda n} .
\end{equation}

In the event that $\delta$ or $\kappa$ becomes large, the Error can similarly
be expanded asymptotically in $\delta+\kappa=\ell$ to see that
\begin{equation}
e^{\hbox{Error}} = 
1 + \Sfrac{\zeta}{4\lambda^2 n^2} 
- \Sfrac{\zeta(4\lambda^2n^2+\zeta)}{(\delta+\kappa)^2\lambda^2n^3}
+ \Sfrac{\zeta c (4\lambda^2n^2+\zeta)}{(\delta+\kappa)^3 \lambda^2n^{7/2}}
+ O((\delta+\kappa)^{-4}) .
\end{equation}
Thus, it follows that, for any fixed $\lambda>0$,
\begin{eqnarray}
&\hspace{-2cm}
A = \frac{
e^{-
(8\lambda(\delta+\kappa)^2n^{3/2}+16c\lambda(\delta+\kappa)n
    -(4(\delta+\kappa)^2-\lambda)\sqrt{n}-8c (\delta +\kappa))
/(8\lambda^2 n^{3/2})
}
}{
\sqrt{\pi\lambda n}
}
\,4^{\lambda n} \cr
& \hspace{3cm} \times \LB 1 + \Sfrac{\zeta}{4\lambda^2 n^2} 
- \Sfrac{\zeta(4\lambda^2n^2+\zeta)}{(\delta+\kappa)^2\lambda^2n^3}
+ O((\delta+\kappa)^{-3} n^{-3/2}) \RB .
\label{69}
\end{eqnarray}
Notice that since $\zeta$ is not a function of $(\delta,\kappa)$ and approaches
a constant as $n\to\infty$, there is a uniform bound on the Error 
for any fixed $\kappa>0$ and $\lambda>0$ when $n\geq N_0>0$ 
and for all $\delta\geq \delta_0>0$ (where $\delta_0$ is a fixed constant).  
In the limit $n\to\infty$ the Error approaches $0$.

Putting $c=1$, $\lambda=\sigma$, $\delta=\omega$ and $\kappa=\delta$ gives
equation \Ref{19} after simplification.  Putting $c=0$ instead, and
$\lambda=\sigma$, $\delta=\omega$ and $\kappa=-\delta$ gives equation
\Ref{18} after simplification.

%

\vspace{1cm}
\noindent{\bf Acknowledgements:} EJJvR acknowledges financial support 
from NSERC (Canada) in the form of Discovery Grant RGPIN-2019-06303.

\vspace{2cm}
\noindent{\bf References}
\bibliographystyle{plain}
\bibliography{trajectory}

\end{document}